\documentclass[aps,pre,twocolumn]{revtex4-1}

\usepackage{amsmath}
\usepackage{amsthm,hyperref}

\usepackage{bbm,graphicx,bm}   
\usepackage[caption=false]{subfig}
\usepackage{color}

\makeatletter
\newcommand*{\defeq}{\mathrel{\rlap{%
                     \raisebox{0.3ex}{$\m@th\cdot$}}%
                     \raisebox{-0.3ex}{$\m@th\cdot$}}%
                     =}
\makeatother

\def\ii{{\rm i}}
\def\sx{\sigma^{\rm x}}
\def\sy{\sigma^{\rm y}}
\def\sz{\sigma^{\rm z}}
\def\tr#1{{\rm tr}(#1)}
\def\1{\mathbbm{1}}
\def\cL{{\cal L}}
\def\cLd{{\cal L}_{\rm dis}}
\def\cLH{{\cal L}_{H}}

\def\ket#1{{| #1 \rangle}}
\def\bra#1{{\langle #1 |}}
\newcommand{\bracket}[3]{{\langle #1 | #2 | #3 \rangle}}
\def\kket#1{{| #1 \rangle\!\rangle}}

\def\tit#1{{\em #1},}
\def\etal#1{#1}

\usepackage{booktabs}
\usepackage{multirow}
\AtBeginDocument{
\heavyrulewidth=.08em
\lightrulewidth=.05em
\cmidrulewidth=.03em
\belowrulesep=.65ex
\belowbottomsep=0pt
\aboverulesep=.4ex
\abovetopsep=0pt
\cmidrulesep=\doublerulesep
\cmidrulekern=.5em
\defaultaddspace=.5em
}

\begin{document}

\title{Relaxation times of dissipative many-body quantum systems}

\author{Marko \v Znidari\v c}
\affiliation{Physics Department, Faculty of Mathematics and Physics, University of Ljubljana, Ljubljana, Slovenia}

\date{\today}

\begin{abstract}
We study relaxation times, also called mixing times, of quantum many-body systems described by a Lindblad master equation. We in particular study the scaling of the spectral gap with the system length, the so-called dynamical exponent, identifying a number of transitions in the scaling. For systems with bulk dissipation we generically observe different scaling for small and for strong dissipation strength, with a critical transition strength going to zero in the thermodynamic limit. We also study a related phase transition in the largest decay mode. For systems with only boundary dissipation we show a generic bound that the gap can not be larger than $\sim 1/L$. In integrable systems with boundary dissipation one typically observes scaling $\sim 1/L^3$, while in chaotic ones one can have faster relaxation with the gap scaling as $\sim 1/L$ and thus saturating the generic bound. We also observe transition from exponential to algebraic gap in systems with localized modes.
\end{abstract}

\pacs{03.65.Yz, 03.65.Aa, 05.70.Ln, 05.30.-d}

\maketitle

\section{Introduction}

With advancing quantum technologies~\cite{NC} it is becoming increasingly important to understand interaction of quantum systems with external degrees of freedom. Evolution of a system coupled to environment can be described by a master equation. A particularly appealing type of a master equation is a Lindblad equation -- a rather general setting that can describe any Markovian evolution~\cite{Breuer}. While Lindblad equations have been used extensively in the past to describe few-particle systems in NMR or quantum optics, recently increasing efforts are devoted to understand many-body systems in the Lindblad setting. Motivation comes from a wide range of fields where Lindblad equations find their application, to name a few: as a computational resource in quantum information, to study e.g. transport properties of strongly-correlated condensed-matter systems, or to study nonequilibrium statistical physics of many-body systems.

Usually the object of most interest for open systems is a steady state, that is, a state to which any initial state converges after a long time. Besides the steady state, dynamics is also of interest with one of the most important quantities being the relaxation time. In a finite system relaxation time is simply equal to the inverse gap of the Liouvillian propagator generating evolution. In statistical physics the scaling power of the gap is called the dynamical exponent -- a critical exponent determining a universality class to which a model belongs. Dynamical exponents have been extensively studied in classical exclusion models, see e.g. Ref.~\cite{Schutz} or Refs.~\cite{Gier06,Popkov} for some more recent results. 

In a quantum domain much less is known about relaxation times of many-body systems. Depending on a situation one might want the gap to be large or small. For instance, if dissipation is engineered in order to prepare a specific steady state one might want relaxation to be as fast as possible. On the other hand, if dissipation is unwanted, for instance, in a quantum memory device, relaxation should be as slow as possible. The value of the gap $g$ is important not just for the relaxation time itself~\cite{Temme10,Wolf15}, but can also carry information about the steady-state properties. Namely, if the gap is finite (so-called rapidly mixing systems) one can show that this implies a clustering of correlations in the steady state~\cite{Poulin10,Nacht11}, meaning that local observables are uncorrelated on a scale larger than $\sim 1/g$. Rapid mixing also implies the stability of steady state to local perturbations~\cite{Lucia13,Kastoryano13,Lucia15}. If the gap on the other hand closes in the thermodynamic limit this can lead to a nonequilibrium phase transition~\cite{Bojan10,PRE11,Kessler12,Fleischauer12,Horstmann13,Bianchi14} and can result in a non-exponential relaxation~\cite{Cai:13,Medvedyeva14} towards a steady state. Understanding how the gap scales with the system size is therefore of fundamental importance.

There have been few scattered results in the literature calculating or bounding the gap either analytically or numerically for specific Lindblad models. With our work we plan to extend these results, studying in more detail how the gap scales with the system size. One can distinguish grossly two different situations depending on the number of sites at which dissipation acts: in a lattice with $L$ sites dissipation acts on $\sim L$ sites (i.e., on most sites), a setting we will call bulk dissipation, or, it can act only on a fixed number of sites (number of sites does not grow with $L$), a setting we will call boundary dissipation (eventhough sites at which it acts need not be at a boundary). Not surprisingly, as we shall see the gap can behave differently in the two cases. What is known so far is that for so-called Davies generators, also called thermal reservoirs, that are an example of bulk dissipation, one is in certain cases able to rigorously prove that the gap is independent of the system length~\cite{Alicki09,Temme13}, $g \sim L^0$. One can also show exponential relaxation in weakly coupled systems~\cite{Schutz15}. Similarly, one can show that the gap can be constant for some other systems with bulk dissipation~\cite{Horstmann13,Cai:13}, while in other systems it can also scale as $\sim 1/L^2$~\cite{troyer05,PRE11,Cai:13}. On the other hand, for open systems with boundary dissipation the observed gaps have so-far all been scaling as $\sim 1/L^3$, or smaller, examples being the XY~\cite{Prosen08,Pizorn08,Medvedyeva14} or the XXX~\cite{JSTAT11,Buca12} model. For scaling of gaps in the Redfield equation see Ref.~\cite{Bojan10}.

In the present work we are going to study the scaling of the gap with system size in a number of spin chain models with bulk as well as with boundary dissipation. The aim is to get a better overview of possible gap scaling and changes in the scaling power as one varies parameters. Such changes can be associated with possible phase transitions in the steady state as well as in decay modes. In the main part of the paper we shall organize sections according to different models studied, explaining for each different techniques used (ranging from analytical to numerical) to infer the scaling. We shall also pay attention to the validity of a weak-dissipation perturbation theory that can be used to calculate the gap, demonstrating its failure in a number of cases. If one is not interested in all the details of gap calculation, or just wants an overview of different gap scalings found, there are Tables~\ref{tab:bulk} and~\ref{tab:bound} provided in the Summary section at the end of the paper.  

The paper is organized as follows. In Sec.II. we briefly explain the setting of Lindblad equations. In Sec III. we then study systems with bulk dissipation, while in Sec. IV. we study systems with boundary dissipation. Each of these two sections is split into subsections describing different systems. In Sec.IV. we also explain an argument that the gap in systems with boundary dissipation can not be larger than $\sim 1/L$. Finally, in Sec.V. we conclude as well as present a summary of the results found.

\section{Lindblad equation}

The Lindblad equation is~\cite{GKS,Lindblad}
\begin{eqnarray}
\frac{{\rm d}\rho}{{\rm d}t}&=&{\cal L}(\rho):=\ii[\rho,H]+\cLd(\rho),
\label{eq:Lin}
\end{eqnarray}
where $\cLd(\rho)=\sum_j 2L_j \rho L_j^\dagger-\rho L_j^\dagger L_j-L_j^\dagger L_j \rho$ is a linear superoperator called a dissipator that can be expressed in terms of traceless Lindblad operators $L_j$. Denoting eigenvalues of the Liouvillian $\cL$ by $\lambda_j(\cL)$, and ordering them according to their real parts, with $\lambda_0(\cL)=0$ assuming to be nondegenerate, the gap is equal to a negative real part of the 2nd largest eigenvalue,
\begin{equation}
g:=-\operatorname{\mathbbm{R}e}[\lambda_1(\cL)]
\end{equation}
Eigenvectors of $\cL$ corresponding to nonzero eigenvalues are called decay modes.

The models that we are going to study will all be spin chains composed of $L$ lattice sites, each carrying a spin-$1/2$ particle. Coupling in $H$ will always be local between nearest-neighbor sites only, and with open boundary conditions. Dissipation will also be local, i.e., each Lindblad operator $L_j$ will act nontrivially only on a single site (different $L_j$ though can act on different sites). Two types of dissipation will be employed, both physically motivated. The first one is dephasing for which $L_j \propto \sz_j$, and which tries to destroy off-diagonal matrix elements (in the eigenbasis of $\sz$). The second one is magnetization driving in which Lindblad operators proportional to raising and lowering operators try to impose an imbalance in populations of spin-up and spin-down states.
 
We mention that the many-body spin chain models studied here are within reach of present day cold-atom technology, e.g. Refs.~\cite{Zoller12,Schwager13}, with individual components like few qubit controlled dissipation~\cite{Barreiro} or Heisenberg spin chains~\cite{Greiner,Bloch} already demonstrated.

\section{Bulk dissipation} 

A canonical model that we shall study in both bulk- and boundary-driven cases is the anisotropic Heisenberg model (XXZ model for short). For zero anisotropy the XXZ model goes into the XX model, which is especially simple and even allows for an exact asymptotic solution.

\subsection{XX with dephasing}

The Hamiltonian of the XX model is
\begin{equation}
H=\sum_{j=1}^{L-1} \sx_j \sx_{j+1}+\sy_j \sy_{j+1}.
\label{eq:XX}
\end{equation}
Dissipation $\cLd$ is given by dephasing of strength $\gamma$ acting independently on each site, which is described by a set of $L$ Lindblad operators,
\begin{equation}
L_j=\sqrt{\frac{\gamma}{2}}\sz_j. 
\label{eq:dephasing}
\end{equation}

We note that such Liouvillian $\cL$ conserves the total magnetization $Z=\sum_j \sz_j$, that is, if $\ket{\psi}$ is an eigenstate of $Z$ with eigenvalue $Z_\psi$ and similarly $\ket{\varphi}$ is an eigenstate of $Z$ with an eigenvalue $Z_\varphi$, then $\cL(\ket{\psi}\bra{\varphi})=\sum_{jk} c_{jk} \ket{\psi_j}\bra{\varphi_k}$ is a superposition of terms in which all $\psi_j$ and all $\varphi_k$ are again eigenstates of $Z$ with eigenvalues $Z_\psi$ and $Z_\varphi$, respectively. More formally conservation means that $U\cL(\rho) U^\dagger=\cL(U\rho U^\dagger)$, with $U$ being rotation around the $z$-axis. $\cL$ therefore has a block structure and we shall label each block by a magnetization difference $z$ and by a number of flipped spins $r$,
\begin{equation}
z:=Z_\psi-Z_\varphi, \quad r:=(L-Z_\psi)/2.
\label{eq:const}
\end{equation}
For an $L$-site chain the allowed values of $z$ are from $-2L$ to $+2L$ in steps of $2$, while that of $r$ are $r=0,1,\ldots,L$. Of special interest will be sectors with $z=0$ because they carry a steady state, i.e., an eigenstate of $\cL$ with eigenvalue $0$. A subspace with $z=0$ and some value of $r$ will be simply called an $r$-particle sector and has (operator) dimension ${L \choose r}^2$. Two most important ones are $r=1$ (1-particle sector), being the smallest nontrivial one, and the largest one with $r=L/2$ (or $r=(L\pm 1)/2$ for odd $L$) being called a half-filling sector because half of the spins are pointing up and half are pointing down.

An important property of dissipative systems where $H$ is quadratic in fermionic operators (via Jordan-Wigner transformation) and Lindblad operators are Hermitian (but not necessarily quadratic), and our XX model is an example of such a system, is that equations for correlation functions split into a hierarchy of equations according to their order, i.e., the number of fermionic operators in the expectation value~\cite{eisler11,Bojan14}. This enables one to exactly calculate few-point expectation values in the steady state, for instance in the presence of an additional boundary driving~\cite{JSTAT10,PRE11,EPJB13}, an incoherent bulk hopping~\cite{eisler11,temme:12}, or special engineered dissipation~\cite{Bojan14}.

Here we are not interested in the steady state but instead in the Liouvillian gap, in particular in its scaling with the system size $L$. Due to magnetization conservation there are $L+1$ steady states, one in each invariant subspace with $z=0$ and a particular $r$. One can also easily see that such a steady state is a uniform mixture off projectors to all ${L \choose r}$ diagonal basis states in the corresponding $r$-particle subspace, for instance, for $L=3$ in a 1-particle sector ($r=1$, $z=0$; subspace dimension $3^2$) the steady state is $\sim \ket{001}\bra{001}+\ket{010}\bra{010}+\ket{100}\bra{100}$. For more details about such states, for instance their Schmidt spectrum, see Appendix~\ref{app:A} . Subspaces with $z \neq 0$ do not contain any zero eigenvalues (i.e., steady states) because they are orthogonal to the identity operator which is, due to trace preservation, always a left eigenvector of $\cL$ with eigenvalue zero and is therefore non-orthogonal to all steady states. Because we are interested in the gap we can limit our discussion to subspaces with $z=0$ as they are the only ones that contain steady states. Subspaces with $r=0$ and $r=L$ are of dimension $1$ and contain only the steady state and are of no interest to us. All other $L-1$ subspaces contain a nontrivial gap. We shall be in particular interested in the smallest gap out of those, i.e., the global one, giving the slowest relaxation rate in the system.

Because of a hierarchy of correlations we know that the eigenvalues, i.e. decay rates, of $p$-point correlation equations are equal to (some) eigenvalues of the Liouvillian~\cite{Bojan14}, however, one in general does not know whether they also give the gap. Regardless of that they can be used as an upper bound on the global gap. For instance, for a model with an incoherent hopping the relaxation rate of $2$-point correlations scales as $\sim 1/L^2$~\cite{eisler11} meaning that the global gap can not be larger. Similarly, for a boundary-driven system numerical calculation of the global gap also resulted in $\sim 1/L^2$ scaling~\cite{PRE11}. 

By numerically diagonalizing our $\cL$ in each $r$-particle sector we observe that the gap is in fact the same in all sectors. This in particular means that to calculate the global gap it is enough to consider the $1$-particle sector which is of dimension ${L \choose 1}^2 =L^2$ (or its symmetric $(L-1)$-particle partner that has exactly the same eigenvalues). This is due to a free nature of $H$, causing that the spectrum of the 1-particle sector to be contained in higher-$r$ sectors. One could be tempted to conclude that the $1$-particle sector will be rather trivial -- this is certainly true for a $1$-particle sector of dimension $L$ in the Hilbert space of states -- however, here we are dealing with a $1$-particle sector of dimension $L^2$ in the Hilbert space of operators. As we shall see this allows for a rich behavior, among other things for a discontinuous transition in the scaling of the gap from a constant $1/L^0$ for small dephasing strength $\gamma$ to $\sim 1/L^2$ for non-small $\gamma$.

\begin{figure}[!h]
\includegraphics[width=3.3in]{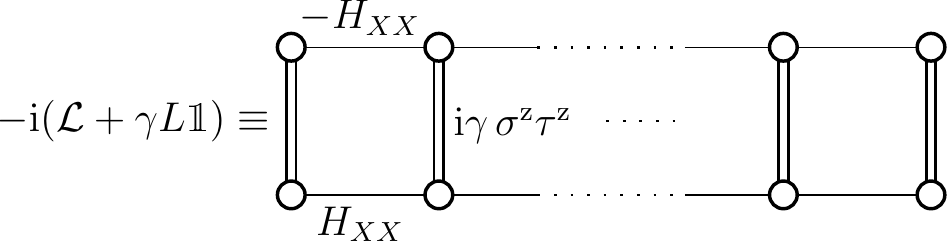}
\caption{The Liouvillian superoperator for the XX chain with dephasing is equivalent to a non-Hermitian ladder Hamiltonian, with $XX$ interaction along upper and lower legs (thin lines) and an imaginary $z-z$ interaction due to dephasing along rungs (double lines).}
\label{fig:lestev}
\end{figure}

Before going to the actual calculation of the gap let us pause for a moment and have a look at an alternative formulation of the eigenvalue problem for the whole $\cL$. The Liouvillian that acts on a $(2^L)^2$ dimensional operator space is a non-Hermitian linear operator that can be written as a ``Hamiltonian'' of a non-Hermitian spin-$1/2$ ladder composed of $L$ rungs. Each rung, spanned by $4$ pure states, takes care of one site of $\cL$ which is also $4$ dimensional (e.g., three Pauli matrices plus an identity). The resulting ladder is for the XX chain with dephasing very simple: it is composed of two $XX$-coupled chains along the ladder legs with an additional imaginary coupling along rungs, see Fig.~\ref{fig:lestev} and Appendix of Ref.~\cite{PRE14} for the mapping details. 
The steady state is an eigenvector corresponding to the eigenvalue zero, in other words a ``ground state'' of a non-Hermitian ladder. Because of dissipative coupling along rungs this ground state is always, regardless of the value of $\gamma$, a direct product of singlet states at each rung. The steady state is therefore always dominated by dephasing $\gamma$, forcing the steady state to be a singlet state. This is due to a very special structure of $\cL$ -- no interaction in $H$ (which plays an inert role) and a dephasing that kills all off-diagonal matrix elements. As we shall show, things are very different for the first decay mode determining the gap. There is a transition from a $\gamma$-dominated phase to a different phase in which $H$ becomes important. 

In the ladder formulation of $\cL$ we can also see why the $1$-particle superoperator sector is nontrivial: it corresponds (in an appropriate basis) to states with one particle in the upper leg (bra) and one particle in the lower leg (ket). Therefore, a $1$-particle superoperator problem is like a problem of $2$ interacting particles on a ladder. It constitutes the simplest case of an interacting system.

\subsubsection{One-particle sector}

Let us now calculate the gap in the 1-particle sector. Essentially the same eigenvalue problem, apart from different boundary conditions~\cite{foot1}, has been rigorously solved in Ref.~\cite{Gaspard:05}. Here we shall present a different and approximate calculation of the gap, which is simpler but nevertheless accurate in the thermodynamic limit $L \to \infty$ in which we are especially interested.

\begin{figure}[t!]
\includegraphics[width=3.3in]{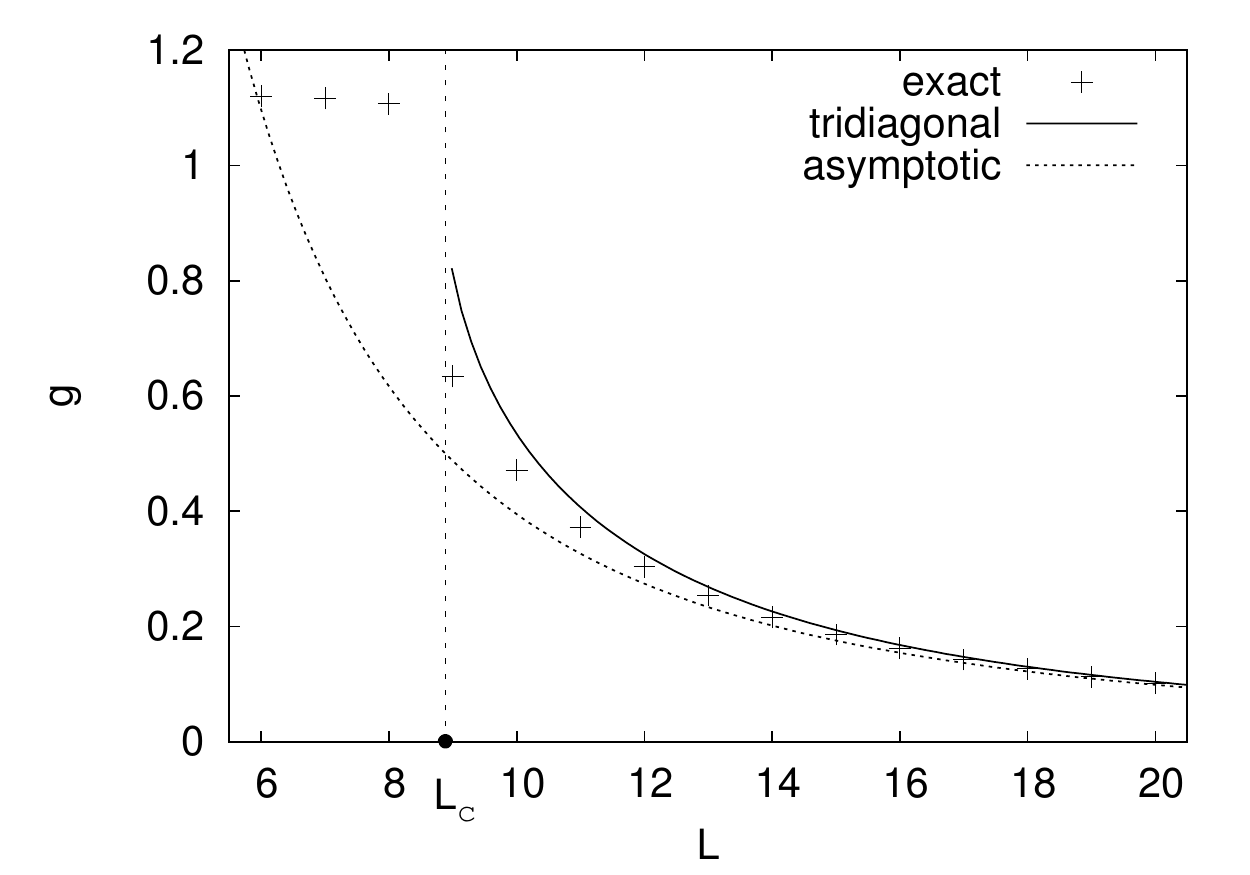}
\caption{The gap $g$ of the XX model with dephasing $\gamma=0.5$. Shown is the exact gap (symbols), the full curve is the approximate formula (\ref{eq:gap1mag}), while the dotted curve shows the asymptotic expression $2\pi^2/(\gamma L^2)$. The approximate result (\ref{eq:gap1mag}) is accurate for $L$ larger than the transition point $L_{\rm c}:=\pi\sqrt{2}/\gamma$.}
\label{fig:gap1mag}
\end{figure}
The basic idea is the following. We have seen that the steady state is always dominated by the dephasing and one can expect that in some range of dephasing strengths the first decay mode will also be of the same nature -- that is dominated by $\gamma$. If this is the case, then its eigenoperator will be close to diagonal because dephasing kills all off-diagonal elements. In the lowest approximation the unitary part $\cL_{\rm H}$ couples diagonal elements $\ket{j}\bra{j}$, where $\ket{j}$ denotes a state of $L$ spins with the $j$-th spin being flipped to $\ket{1}$ while all others are in state $\ket{0}$, to two off-diagonal $\ket{j}\bra{j+1}$ and $\ket{j+1}\bra{j}$. Such a tri-diagonal approximation reduces the size of $\cL$ in the 1-particle sector from $L^2$ to $3L-2$. In addition, we can immediately see that on this three-diagonal subspace all $(L-1)$ states of the form $\ket{j}\bra{j+1}+\ket{j+1}\bra{j}$ are eigenstates with eigenvalue $-4\gamma$ (they are eigenstates of $\cLH$ with eigenvalue $0$ because of different signs when $H$ acts from the left and from the right, while $\cLd(\ket{0}\bra{1}_j)=-2\gamma \ket{0}\bra{1}_j$ and similarly for $\ket{1}\bra{0}_j$). Reduction of $\cL$ to $(2L-1)$ dimensional subspace spanned by $\ket{j}\bra{j}$ and $\ket{j}\bra{j+1}-\ket{j+1}\bra{j}$ has a block structure of form,
\begin{equation}
\cL_{\rm red}=
\left( \begin{array}{cc}
0 & C^{\rm T} \\
C & -4\gamma \mathbbm{1} \\
\end{array} \right), \quad C_{j,k}=-2{\rm i}\sqrt{2}(\delta_{j,k}-\delta_{k,j+1}).
\end{equation}
Basis ordering is such that the first diagonal block corresponds to $L$ states $\ket{j}\bra{j}$, and the second to $(L-1)$ states $\ket{j}\bra{j+1}-\ket{j+1}\bra{j}$. Writing the eigenvector as $(\mathbf{x},\mathbf{y})$ and the eigenvalues as $\lambda$, and eliminating $\mathbf{x}$, we get an eigenvalue equation $C C^{\rm T}\mathbf{y}=(\lambda^2+4\gamma \lambda)\mathbf{y}$, with $C\,C^{\rm T}$ being a tridiagonal matrix of size $(L-1)\times (L-1)$ with matrix elements
\begin{equation}
(CC^{\rm T})_{j,k}=8(-2\delta_{j,k}+\delta_{j+1,k}+\delta_{j-1,k}).
\end{equation}
$CC^{\rm T}$ is nothing but a standard matrix appearing is a solution of harmonic oscillators or a tight-binding model, and has eigenvalues $-32\cos^2{(\pi j/(2L))}, \quad j=1,\ldots L-1$, leading to eigenvalues $\lambda_j$ of $\cL_{\rm red}$ being $\lambda_j=-2\gamma(1\pm \sqrt{1-8\cos^2{(\pi j /(2L))}/\gamma^2})$. The gap is determined by the largest eigenvalue $\lambda_{j=L-1}$, and is
\begin{equation}
g=2\gamma\left(1-\sqrt{1-\frac{8}{\gamma^2}\sin^2{(\frac{\pi}{2L})}}\right) \asymp \frac{2\pi^2}{\gamma L^2}+\cdots.
\label{eq:gap1mag}
\end{equation}
The gap $g$ gives the distance of the largest decay mode eigenvalue from the origin along a real axis, with the next largest decay mode being asymptotically at a distance $4g$. For $H$ with periodic boundary conditions the gap is $4$ times larger than the above $g$ (\ref{eq:gap1mag}). We can see that for sufficiently small $\gamma$ the expression under the square root in Eq.(\ref{eq:gap1mag}) can become negative. For large $L$ this happens for $\gamma < \gamma_{\rm c}$, where the critical dephasing is 
\begin{equation}
\gamma_{\rm c} = \frac{\pi \sqrt{2}}{L}.
\label{eq:gammac}
\end{equation}
\begin{figure}[t!]
\includegraphics[width=3.3in]{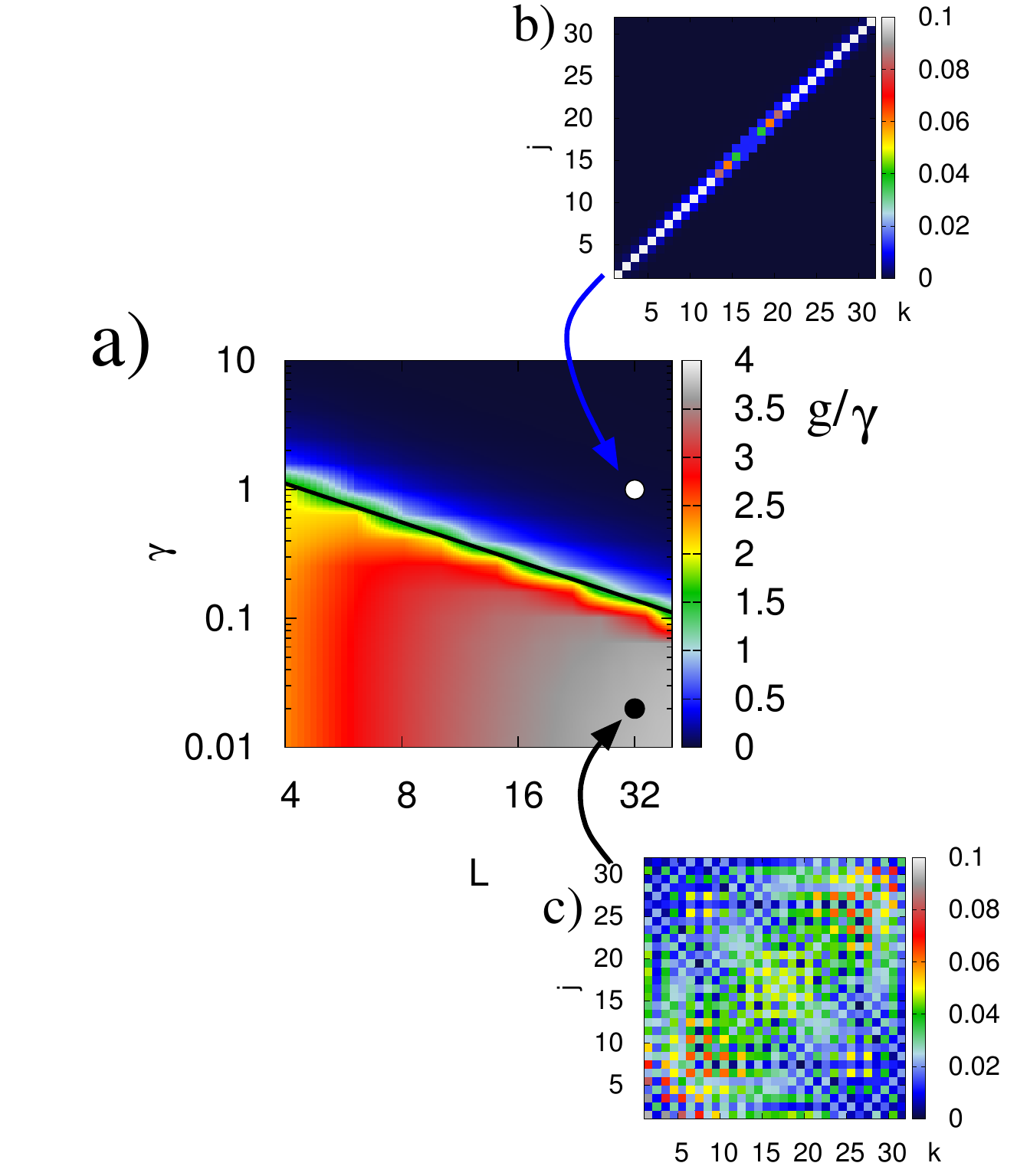}
\caption{(Color online) Phase diagram of the largest decay mode. (a) Dependence of gap $g/\gamma$ on dephasing and $L$. Full black line is the critical $\gamma_{\rm c}$ (\ref{eq:gammac}) delimiting two phases. In (b) and (c) are shown expansion coefficients $|c_{k,j}|$ of the largest decay mode $x$, $x=\sum_{k,j=1}^L c_{k,j} \ket{k}\bra{j}$, showing an almost diagonal dephasing-dominated phase with $g \sim 1/L^2$ in (b), and a delocalized phase with $g \sim 1/L^0$ in (c).}
\label{fig:gap1mag_faze}
\end{figure}
Our approximate expression for the gap (\ref{eq:gap1mag}) is accurate only for $\gamma \gg \gamma_{\rm c}$, where the decay mode is indeed governed by dephasing, see Fig.~\ref{fig:gap1mag} . Note though that in the thermodynamic limit $\gamma_{\rm c} \to 0$ and therefore Eq.~(\ref{eq:gap1mag}) becomes exact for any $\gamma$. One can see already in Fig.~\ref{fig:gap1mag} that below $\gamma_{\rm c}$ (or $L_{\rm c}$, depending on which parameter is held fixed) the gap becomes independent of $L$ and in the thermodynamic limit equal to $4\gamma$ (in this regime the largest decay mode eigenvalue is complex and the gap $g$ is equal to its real part). Therefore, for the largest decay mode there are two phases: for $\gamma \ll \gamma_{\rm c}$ the decay mode is governed by $H$ with the gap being $g = 4\gamma$, while for $\gamma \gg \gamma_{\rm c}$ the decay mode is governed by dephasing and the gap is $g \asymp \frac{2\pi^2}{\gamma L^2}$. One can use perturbation theory to show that the gap is indeed independent of system size for sufficiently small dephasing~\cite{Horstmann13}. The convergence radius of weak-coupling perturbation series for small $\gamma$ is $\gamma_{\rm c}$ (\ref{eq:gammac}) and algebraically shrinks to zero in the thermodynamic limit. Also, the thermodynamic limit $L \to \infty$ and the weak coupling limit $\gamma \to 0$ do no commute, which can be also seen in Fig.~\ref{fig:gap1mag_faze} showing the phase diagram.

\begin{figure}[t!]
\includegraphics[width=3.3in]{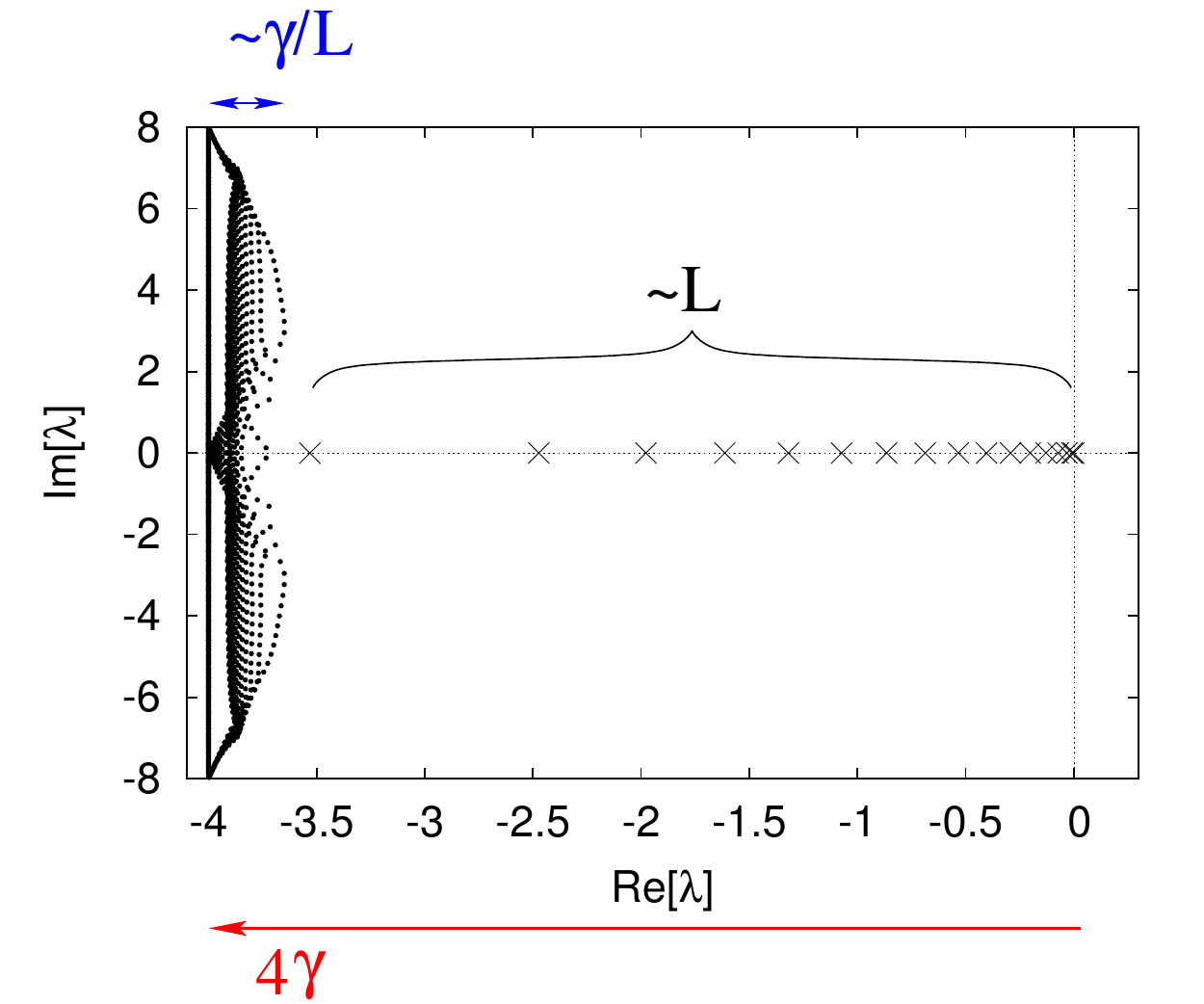}
\caption{(Color online) Complex eigenvalues of $\cL$ for XX model with bulk dephasing, $\gamma=1$, $L=50$, and 1-particle sector. For $\gamma > \gamma_{\rm c}$ there are of order $\sim L$ eigenvalues that are on the real axis and are separated from the bulk consisting of the remaining $\sim L^2$ complex eigenvalues. Real parts of the bulk eigenvalues are around $-4\gamma$ with the width being $\sim \gamma/L$.}
\label{fig:eig1mag}
\end{figure}
Finally, let us briefly comment also on the overall spectrum of $\cL$. Numerically diagonalizing $\cL$ on the 1-particle subspace the following picture emerges, see Fig.~\ref{fig:eig1mag} . Most of the eigenvalues, $\sim L^2$ in number, are within the bulk laying in a complex plane. As one increases $L$ eigenvalues ``evaporate'' from the bulk and join a bunch of real eigenvalues to the right of the bulk. The number of these separated real eigenvalues is proportional to $L$. Decay modes corresponding to the separated real eigenvalues are dictated by dephasing, whereas decay modes in the bulk are instead dictated by the Hamiltonian. Remember that in the absence of dephasing the spectrum of $\cL$ would be entirely on the imaginary axis. Such distinct nature of two decay mode types is in turn reflected in two different phases of the largest decay mode, Fig.~\ref{fig:gap1mag_faze} . Indeed, if one decreases $\gamma$ at fixed $L$, the real eigenvalues get ``absorbed'' in the bulk, with the last one disappearing at $\approx \gamma_{\rm c}$, at which point a transition happens in the scaling of the gap with $L$. Approximate values of real eigenvalues can be obtained from the approximation with $\cL_{\rm red}$ (\ref{eq:gap1mag}), resulting in a scaling of the $j$-th largest eigenvalue of $\cL$ as $\lambda_j \sim (j/L)^2$. In the thermodynamic limit real eigenvalues therefore cluster around the origin with their density there having a square-root singularity. It has been observed~\cite{Cai:13} that if the gap $g$ closes in the thermodynamic limit there is a possibility for an algebraic relaxation instead of an exponential one occurring when $g$ is finite. Such an algebraic decay can be explained~\cite{Medvedyeva14} by clustering of eigenvalues around $0$. In our XX chain with dephasing this clustering is of the same kind as in boundary-driven free models studied in Ref.~\cite{Medvedyeva14} and as a consequence, in the thermodynamic limit the relaxation will have a power-law form.

\subsection{XXZ with dephasing}

In the XX model the bulk Hamiltonian describes non-interacting particles. Choosing the XXZ Hamiltonian instead, the additional coupling in the $z$-direction represents the interaction between fermions in the Jordan-Wigner picture. In this section we shall therefore consider the XXZ spin chain with bulk dephasing. The Hamiltonian of the XXZ chain is
\begin{equation}
H=\sum_{j=1}^{L-1} \sx_j \sx_{j+1}+\sy_j \sy_{j+1}+\Delta \sz_j \sz_{j+1},
\label{eq:XXZ}
\end{equation}
with $\Delta$ being an anisotropy parameter. Dissipation is the same dephasing at each site as used already for the XX model (\ref{eq:dephasing}). Magnetization is again conserved and there is one steady state in each sector with magnetization difference $z=0$ and $r$ particles, see discussion for the XX model with dephasing. Using a Jordan-Wigner transformation the anisotropy part $\Delta \sz_j \sz_{j+1}$ has a form $\sim \Delta n_j n_{j+1}$ and therefore represents interaction between nearest-neighbor fermions. Analytical solution for eigenvalues of $\cL$ is not possible anymore and we will have to resort to various perturbation approaches and numerical calculation.

Numerically calculating the gap one sees that, as opposed to the XX model, the gap is this time different in different sectors. In particular, the global, i.e., the smallest gap, is from the half-filling sector with $r=L/2$ (or $r=(L\pm 1)/2$ for odd $L$) and not from the 1-particle sector, $r=1$. We shall nevertheless first discuss the 1-particle sector, where in the thermodynamic limit things are simple.

\subsubsection{One-particle sector}
\begin{figure}[t!]
\includegraphics[width=3.3in]{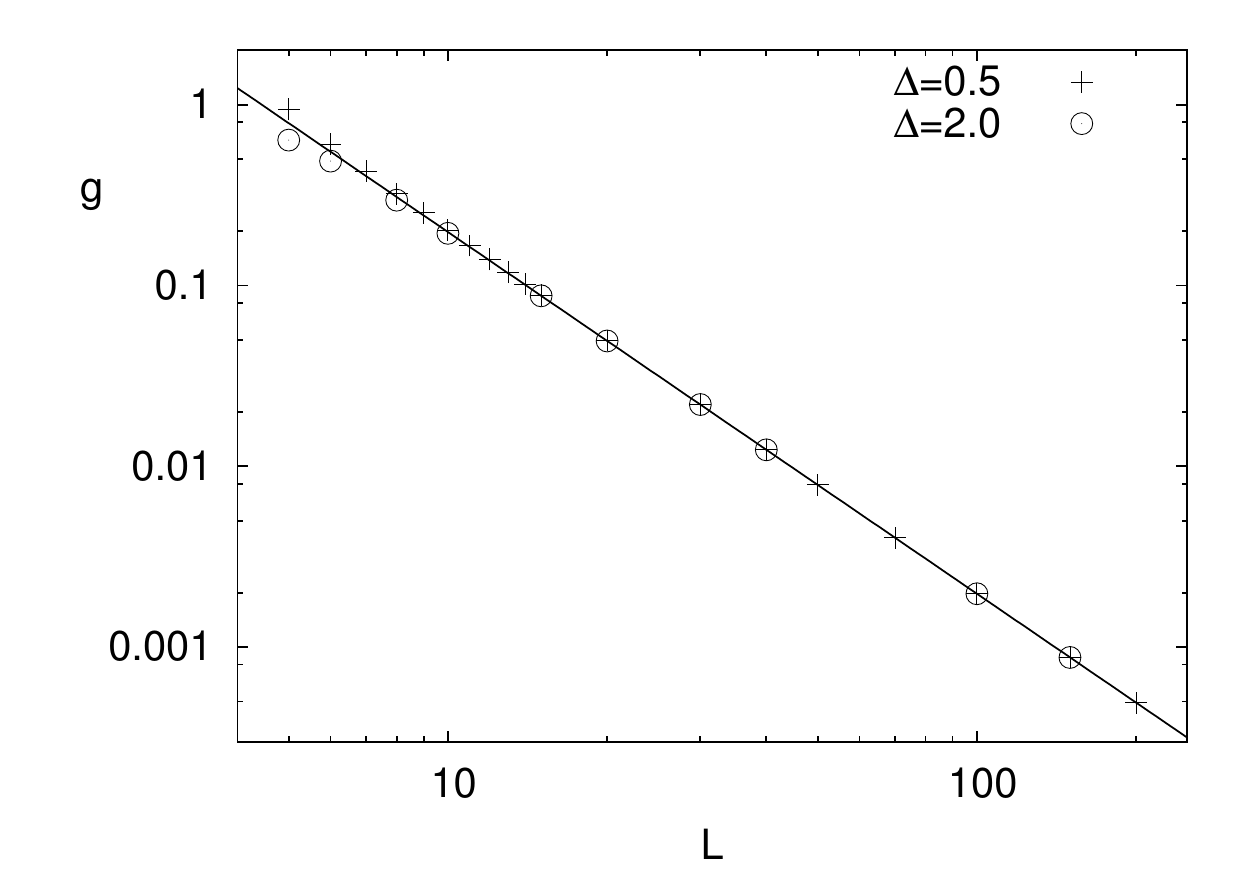}
\caption{The gap $g$ in the 1-particle sector of the XXZ chain with dephasing, $\gamma=1$. Symbols show results of numerical diagonalization for anisotropies $\Delta=0.5$ and $2.0$, while the full line is $2\pi^2/L^2$, giving the asymptotic gap, which is equal to the one in the XX chain with dephasing (\ref{eq:gap1mag}).}
\label{fig:XXZ-1mag}
\end{figure}

Because the 1-particle sector's size is $L^2$ one can numerically calculate the gap for systems of reasonable size. Results are in Fig.~\ref{fig:XXZ-1mag} .
As we can see for large system sizes the gap is equal to
\begin{equation}
g \asymp \frac{2\pi^2}{\gamma L^2},
\label{eq:XXZ_1mag}
\end{equation}
and is therefore equal to the one for the XX chain with dephasing. In the 1-particle sector the gap is in the thermodynamic limit independent of interaction (anisotropy) $\Delta$. This can be qualitatively understood if one looks at the non-Hermitian ladder formulation of $\cL$ (for the XXZ chain the only difference compared to the XX chain in Fig.~\ref{fig:lestev} is that the Hamiltonian along the two legs is the XXZ Hamiltonian, see Ref.~\cite{PRE14}). Because there are only two particles present in the ladder interaction is infrequent, it is in fact only a $\sim 1/L$ boundary effect~\cite{foot2}, and is asymptotically irrelevant (for finite $L$ correction to the asymptotic gap (\ref{eq:XXZ_1mag}) due to finite $\Delta$ scales as $\sim 1/L^4$).

\subsubsection{Global gap}

As mentioned, for the XXZ chain with dephasing the smallest gap is from the so-called half-filling sector with $r=L/2$ particles, which is also the largest subspace of $\cL$. Here we shall study the gap in this sector. We are going to use perturbation theory for small $\gamma$ as well as for large $\gamma$, while in-between numerical calculations will be used to infer a general form of $g$. We are also going to show that the perturbation theory for large $\Delta$ fails.

Let us start with small $\gamma$. For the unperturbed part of the Liouvillian we take the whole unitary part, $\cL_0:=\ii[\rho,H]$, generated by the XXZ Hamiltonian (\ref{eq:XXZ}), while perturbation is the dephasing $\cL_1:=\cLd$. Because the unperturbed $\cL_0$ has a degenerate steady state subspace corresponding to eigenvalue $0$, we have to use 1st order degenerate perturbation theory. The steady-state subspace is spanned by all projectors $x_k=\ket{\psi_k}\bra{\psi_k}$ to eigenstates $\ket{\psi_k}$ of $H$ (we assume a non-degenerate spectrum of $H$). Let us denote a projection of $\cL_1$ to the steady-state subspace of $\cL_0$ by $\gamma R:={\cal P}\cL_1{\cal P}$, with $R_{j,k}=\tr{x_j \cL_1(x_k)}$ being of size ${L \choose r}$. Using the form of dephasing Lindblad operators (\ref{eq:dephasing}) we get
\begin{equation}
R_{j,k}=-L \delta_{j,k}+\sum_{p=1}^L |\bracket{\psi_j}{\sz_p}{\psi_k}|^2.
\label{eq:R}
\end{equation}
The largest eigenvalue of $R$ is equal to $0$ and the next-largest one, denoted by $-c_1$ (similarly as for $\cL$, all eigenvalues of $R$ have non-positive real parts), determines the gap of $\cL$ for small $\gamma$, 
\begin{equation}
g = \gamma\, c_1 + {\cal O}(\gamma^2).
\label{eq:XXZ-weak}
\end{equation}
We observe that, while on one hand dephasing can be derived as being due to a classical fluctuating magnetic field in the $z$-direction~\cite{Carmichael}, we also see that the gap of $\cL$ for small dephasing is determined by eigenstate fluctuations of magnetization in the $z$-direction, as reflected in the form of $R$ (\ref{eq:R}).
\begin{figure}[t!]
\includegraphics[width=3.3in]{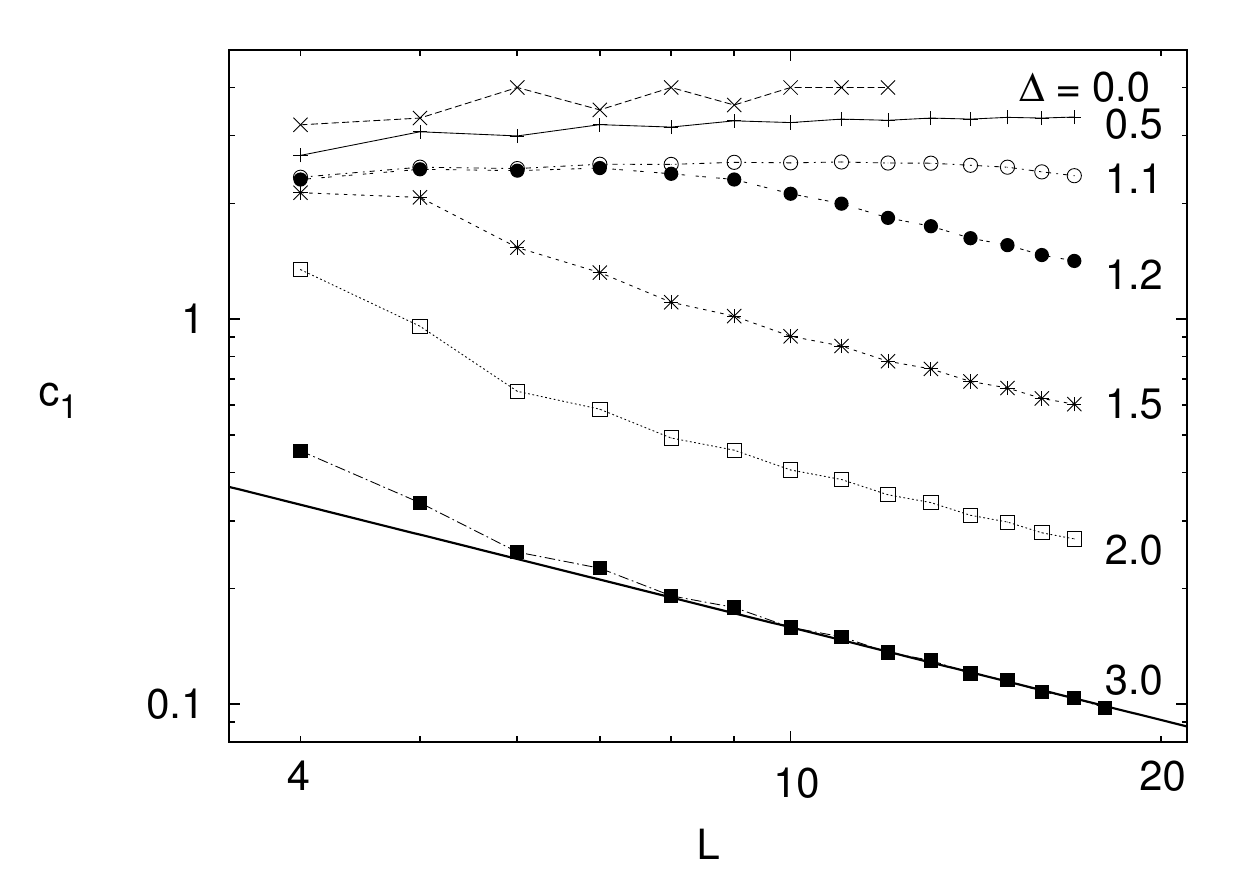}
\caption{The largest non-zero eigenvalue $c_1$ of matrix $R$ (\ref{eq:R}) in the XXZ chain with dephasing, determining the gap Eq.~(\ref{eq:XXZ-weak}). Full line is $1/L^{0.8}$, suggesting the asymptotic dependence for $\Delta  >1$.}
\label{fig:XXZ-weak}
\end{figure}
In Fig.~\ref{fig:XXZ-weak} we show results of exact numerical diagonalization of $R$ for different anisotropies and system sizes, all in the largest half-filling sector with $r=L/2$. We can see that for $\Delta<1$ the eigenvalue $c_1$ becomes asymptotically independent of $L$, for instance, for $\Delta=0$ it converges towards $4$ as already calculated in section about the XX model. For $\Delta >1$ it on the other hand decays; numerical data for available $L\le 18$ is consistent with a $\sim 1/L^{0.8}$ decay. For small $(\Delta-1)$ this asymptotic decay starts for sizes larger than $L_* \propto 1/(\Delta-1)$. It would be interesting if one would be able to directly calculate the largest nontrivial eigenvalue of $R$ via the Bethe ansatz eigenstates $\psi_k$.

So far we have calculated the global gap for small dephasing $\gamma$ without saying anything about the convergence radius, i.e., the validity of such an approximation. To find out critical $\gamma_{\rm c}$ upto which perturbative gap (\ref{eq:XXZ-weak}) can be expected to hold, we are going to numerically calculate the gap and compare it to perturbative prediction.
\begin{figure}[t!]
\centerline{\includegraphics[width=1.62in]{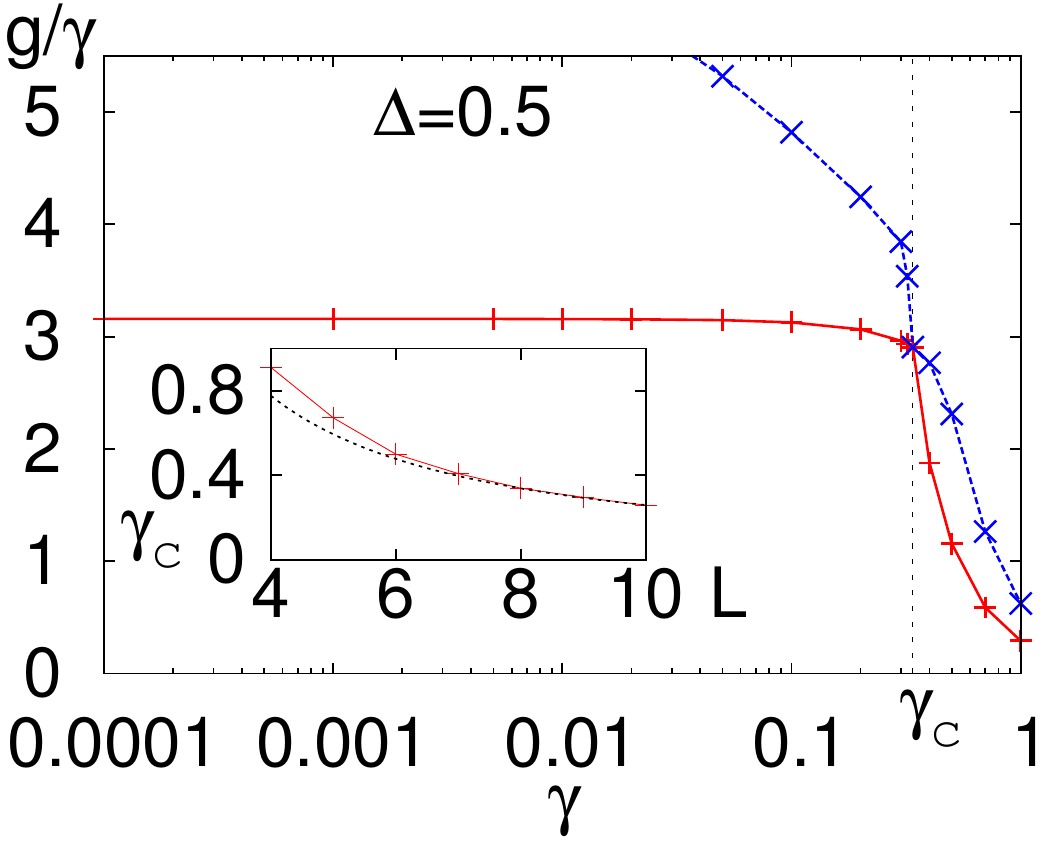}
\includegraphics[width=1.58in]{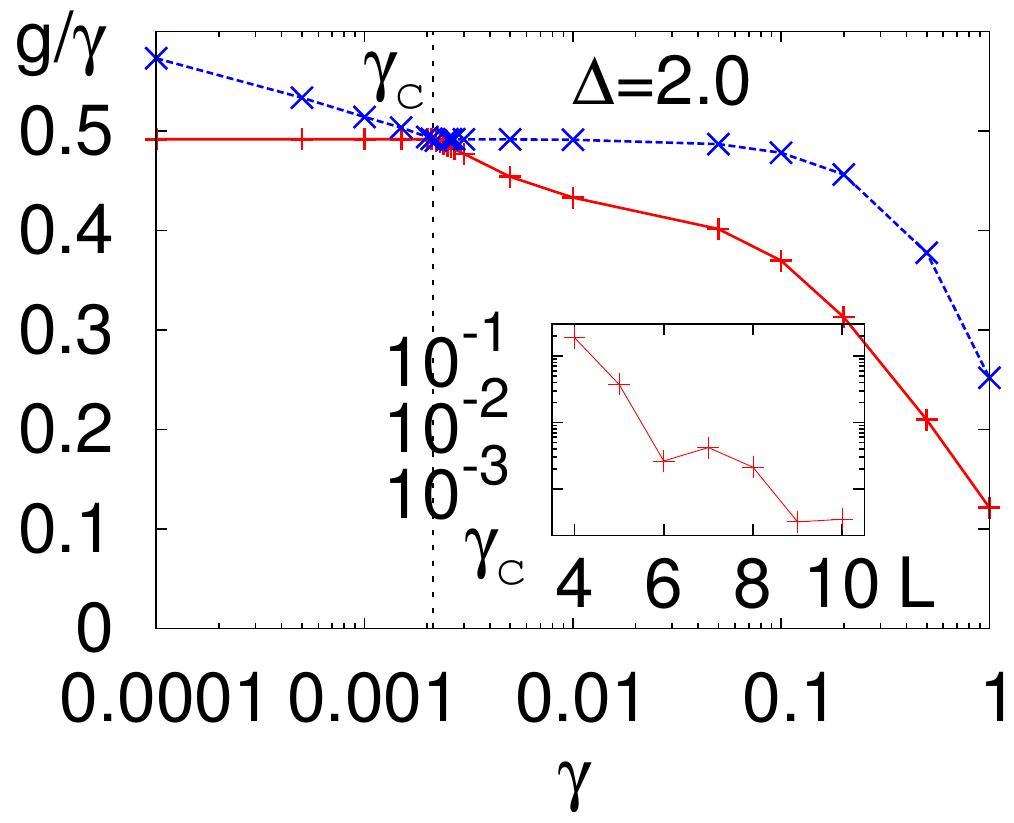}}
\caption{(Color online) The largest non-zero eigenvalue in the even/odd sector for the XXZ chain with dephasing, all for $L=8$. Insets show dependence of the critical $\gamma_{\rm c}$ that determines validity of perturbation series. Left: gapless regime of $\Delta=0.5$ for which $\gamma_{\rm c}$ depends on $L$ algebraically (dashed curve in the inset is $\sim 1/L^{1.2}$). Right: gapped regime $\Delta=2.0$ for which $\gamma_{\rm c} \asymp \exp{(-k L)}$ (inset).}
\label{fig:XXZ-pert}
\end{figure}
In addition to symmetries of $\cL$ already pointed out there is also a spatial reflection symmetry with respect to exchanging sites $j \to L-j$, resulting in even and odd decay modes and eigenvalues of $\cL$. The largest nonzero eigenvalue in even and odd sector (both for $z=0$ and half-filling $r=L/2$) is shown in Fig.~\ref{fig:XXZ-pert} for system size $L=8$ and two values of $\Delta$. The steady state -- a uniform combination of all pure-state projectors -- is always from the even sector. We can see that the largest non-zero eigenvalues cross at a critical $\gamma_{\rm c}$. The decay mode that determines the gap in the half-filling sector (red pluses in Fig.~\ref{fig:XXZ-pert}) is from the even sector for $\gamma < \gamma_{\rm c}$ and from the odd one for $\gamma > \gamma_{\rm c}$. For $\gamma < \gamma_{\rm c}$ we see that the gap (red pluses) is proportional to $\gamma$ (horizontal line in Fig.~\ref{fig:XXZ-pert}) and therefore perturbation series (\ref{eq:XXZ-weak}) holds. The convergence radius $\gamma_{\rm c}$ shrinks algebraically with $L$ in the gapless phase $\Delta <1$, at $\Delta=0.5$ it decays as $\gamma_{\rm c} \asymp 4.1/L^{1.2}$, although we can not exclude asymptotic $\sim 1/L$ scaling, while it is exponentially small in the gapped phase $\Delta >1$. For the XXZ chain with bulk dephasing the convergence radius of perturbation series in the dephasing therefore shrinks to zero in the thermodynamic limit. We also note that the two eigenvalues shown in Fig.~\ref{fig:XXZ-pert} are both real, except the odd one for $\gamma < \gamma_{\rm c}$ and $\Delta=0.5$ which forms a complex pair.

\begin{figure}[t!]
\includegraphics[width=3.3in]{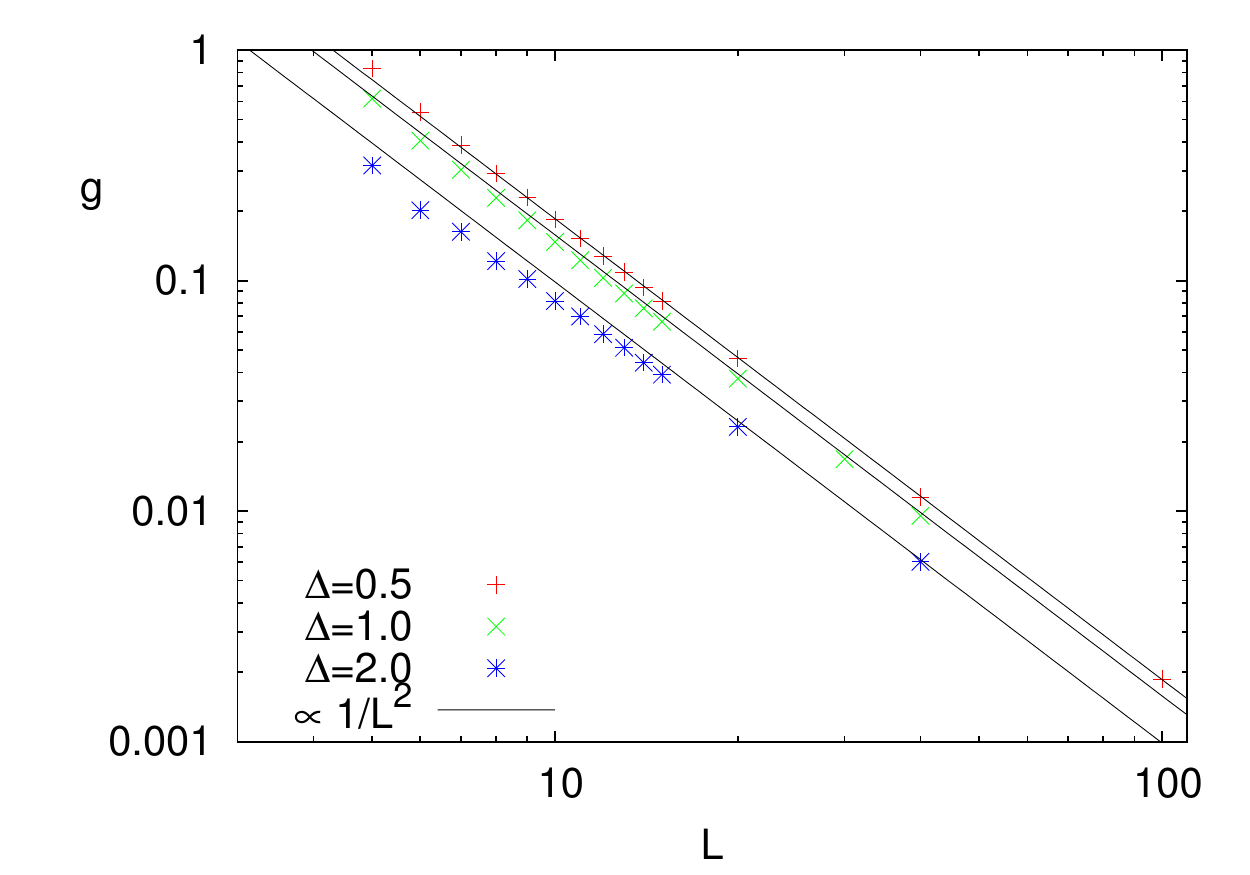}
\caption{(Color online) Gap for the XXZ chain with dephasing in the half-filling sector, $\gamma=1$. Full lines denote asymptotic $\sim 1/L^2$ scaling for all $\Delta$.}
\label{fig:XXZ-gap}
\end{figure}

For non-small $\gamma$ the XXZ chain with bulk dephasing is diffusive~\cite{NJP10} and therefore one in general expects the gap to scale as $\sim 1/L^2$ in the thermodynamic limit. Namely, for diffusive systems evolution of macroscopic observables does not depend independently on time $t$ and spatial coordinate $x$ but instead only on one scaling variable $x^2/t$ and therefore time should scale with a size squared. We numerically calculated~\cite{foot3} the gap for fixed non-small $\gamma$ and indeed confirmed the $g \sim 1/L^2$ scaling, see Fig.~\ref{fig:XXZ-gap} as well as Ref.~\cite{Cai:13}.
Exact dependence of $g$ on $\gamma$ and $\Delta$ is more complicated; a crude approximation that seems to work for large $L$ (at least for $\gamma=1$) is $g \approx \frac{2\pi^2}{\gamma L^2}\frac{4\gamma^2}{4\gamma^2+\Delta^2}$ -- the straight lines plotted in Fig.~\ref{fig:XXZ-gap} is in fact this expression. 

When some parameters in the system are large one can again use perturbation theory. One such possibility is the case of large dephasing $\gamma$, where the unperturbed Liouvillian $\cL_0$ contains $H_0=\sum_j \Delta \sz_j \sz_{j+1}$ and dephasing dissipation $\cLd$ while the perturbation $\cL_1$ consists of hopping given by $H_1=\sum_j \sx_j \sx_{j+1}+\sy_j \sy_{j+1}$. This has been done in Ref.~\cite{Cai:13}, obtaining that the gap for large $\gamma$ scales as $\sim 1/(\gamma L^2)$.

We shall redo calculations of Ref.~\cite{Cai:13}, deriving also subleading terms, in order to be able to comment on the validity of different limits. First, $\cL_0$ has a degenerate steady-state subspace and one has to use degenerate perturbation theory. First-order perturbation on the steady-state subspace of dimension $2^L$ is always zero because all steady states are diagonal, while the hopping $\cL_1$ transports one spin in the bra or ket, and so at least two applications of $\cL_1$ are needed to again get a diagonal state and with it a nonzero matrix elements of $\cL_1$. Second-order perturbation of the steady state manifold is determined by,
\begin{equation}
\cL_{\rm eff}=-{\cal P}\cL_1 \cL_0^{-1} \cL_1 {\cal P},
\label{eq:pert2}
\end{equation}
where ${\cal P}$ is a projection operator to the steady state manifold. The steady state manifold of $\cL_0$ consists of all diagonal density matrices, $\ket{\psi_j}\bra{\psi_j}$, where $\ket{\psi_j},\quad j=1,\ldots 2^L$ are $2^L$ (basis) states in the standard $\sz$-eigenbasis. One can write matrix elements of $\cL_{\rm eff}$ as $[H_{\rm eff}]_{j,k}:=-\tr{\ket{\psi_j}\bra{\psi_j} \cL_{\rm eff}(\ket{\psi_k}\bra{\psi_k})}$, where matrix $H_{\rm eff}$ can in turn be written in terms of Pauli spin variables, obtaining
\begin{eqnarray}
H_{\rm eff}&=& \frac{H_{\rm XXX}}{\gamma} +\frac{\Delta^2/\gamma}{\Delta^2+4\gamma^2}\left[\vec{\sigma}_{1}\cdot \vec{\sigma}_{2}+\vec{\sigma}_{L-1}\cdot \vec{\sigma}_{L} -2\cdot \1 \right]+ \nonumber \\
&&\!\!\!\!\!\!\! +\frac{\Delta^2/\gamma}{2(\Delta^2+\gamma^2)} \sum_{j=1}^{L-3} (\sz_j\sz_{j+3}-\1)(\1-\vec{\sigma}_{j+1}\cdot \vec{\sigma}_{j+2}),
\label{eq:Heff}
\end{eqnarray}
where $H_{\rm XXX}:=\sum_{j=1}^{L-1} (\1 -\vec{\sigma}_{j}\cdot \vec{\sigma}_{j+1})$. Such $H_{\rm eff}$ is Hermitian, has $L+1$ zero eigenvalues (one for each invariant magnetization sector), while all other eigenvalues are positive. The smallest non-zero eigenvalue is equal to the gap $g$ of $\cL$ in the limit when $\cL_1 \ll \cL_0$. In Ref.~\cite{Cai:13} the above result (\ref{eq:Heff}) has been derived in the leading order in $1/\gamma$, i.e., $H_{\rm eff}=H_{\rm XXX}/\gamma$, from which due to a quadratic low-energy dispersion of $H_{\rm XXX}$ one gets the asymptotic gap $g \asymp \frac{2\pi^2}{\gamma L^2}$. This perturbative result is also valid in the thermodynamic limit~\cite{Cai:13} because the convergence radius does not shrink to zero. With the exact expression for $H_{\rm eff}$ we can also explore the case of large $\Delta$, for which one would be tempted to think that perturbative series will again work because $\cL_1$ is again small compared to $\cL_0$ that contains a large parameter $\Delta$. Such reasoning though is in fact wrong and one can not use perturbation in $\cL_1$ if only $\Delta$ is large. The reason for the failure is rather instructive and we are going to explain why it happens. As opposed to the limit $\gamma \to \infty$, for $\Delta \to \infty$ the two terms that appear in addition to $H_{\rm XXX}$ in Eq.~(\ref{eq:Heff}) can not be neglected because the two prefactors in front of them scale for large $\Delta$ as $1/\gamma$ and are therefore of the same order as the $H_{\rm XXX}$ term. There is also an additional subtlety: it would be tempting to retain only the leading order expansion of the two prefactors, the already mentioned $1/\gamma$ and $1/(2\gamma)$, however, in that case the ground state of $H_{\rm eff}$ is highly degenerate and the gap would therefore remain zero (one of the reasons for the degeneracy is that the boundary term in the first line of Eq.~(\ref{eq:Heff}) with a $1/\gamma$ prefactor exactly cancels the boundary terms from $H_{\rm XXX}$, leaving the first and the last spin uncoupled). One has to retain at least the 1st order expansion of the prefactors, resulting in terms that are proportional to $\frac{\gamma}{\Delta^2}$. This means that the gap of $H_{\rm eff}$, and with it also the Liouvillian gap $g$, will scale as $g \propto \frac{\gamma}{\Delta^2}$ for large $\Delta$ (this conclusion remains true despite the failure of 2nd order perturbation expansion).   
\begin{figure}[!h]
\includegraphics[angle=-90,width=3.2in]{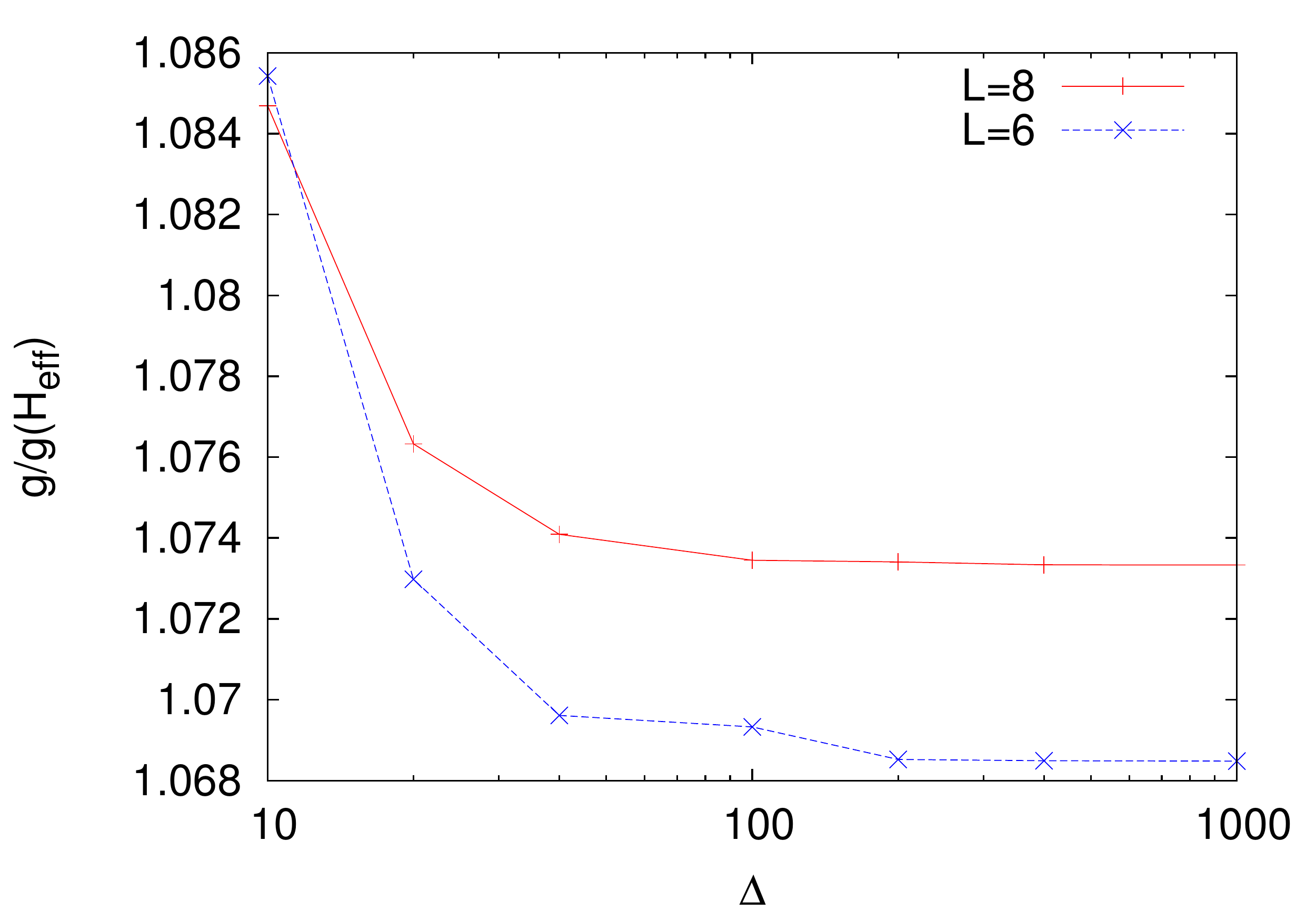}
\caption{(Color online) Relative error of the gap calculated from perturbative $H_{\rm eff}$ (\ref{eq:Heff}), $\gamma=1$. For large $\Delta$ the error does not go to zero, signifying that the second-order perturbation theory in large $\Delta$ fails.}
\label{fig:Heff}
\end{figure}
Calculating the gap of the full $H_{\rm eff}$ (\ref{eq:Heff}) and comparing it with the exact gap of $\cL$, one gets data in Fig.~\ref{fig:Heff} . As one can see the error does not decrease to zero as one increases $\Delta$ (this residual error as well increases with $L$). The reason is that for large $\Delta$ and fixed dissipation $\gamma$ not all non-zero eigenvalues of $\cL_0$ are large -- some are of order $\gamma$ -- and therefore a pseudoinverse $\cL_0^{-1}$ in perturbation series (\ref{eq:pert2}) is not necessarily small. This occurs because the spectrum of $H_0$ is highly degenerate with eigenstates being product states in the eigenbasis of $\sz$, which is also the eigenbasis of $\cLd$. If $\ket{\psi}$ and $\ket{\varphi}$ are two such degenerate eigenstates we will have $[\ket{\psi}\bra{\varphi},H_0]=0$ while for dephasing $||\cLd(\ket{\psi}\bra{\varphi})|| \sim \gamma $. As a consequence, consecutive orders in perturbation series do not necessarily uniformly decrease. This can be compared to perturbation series for fixed $\Delta$ and large $\gamma$, for which all non-zero eigenvalues of $\cL_0$ are large and of order $\gamma$ and therefore $\cL_0^{-1}$ is small. One can in fact draw a general rule: if a large perturbation parameter is in global dissipation (strong coupling) a perturbative series will be well behaved, whereas if a large parameter is only in $H_0$ one must be careful if $H_0$ has degeneracies. Similar complications can occur if (strong) dissipation acts only on a few sites.

Let us summarize our findings for the XXZ chain with bulk dephasing and the half-filling sector: for small dephasing $\gamma < \gamma_{\rm c}$ the gap is $\sim 1/L^0$ for $\Delta <1$, while it scales as $\sim 1/L^{0.8}$ for $\Delta>1$. Critical dephasing $\gamma_{\rm c}$ decays algebraically with $L$ for $\Delta <1$ while it is exponentially small for $\Delta > 1$, see also Table~\ref{tab:bulk} . Perturbation series in small dephasing therefore fails in the thermodynamic limit. For non-small dephasing the gap is $\sim 1/L^2$ regardless of $\Delta$, as one would expect for a diffusive system. Perturbation series for large $\gamma$ works, while it fails if $\Delta$ is the only large parameter.

\subsection{Constant gap}
\begin{figure}[t!]
\includegraphics[width=3.2in]{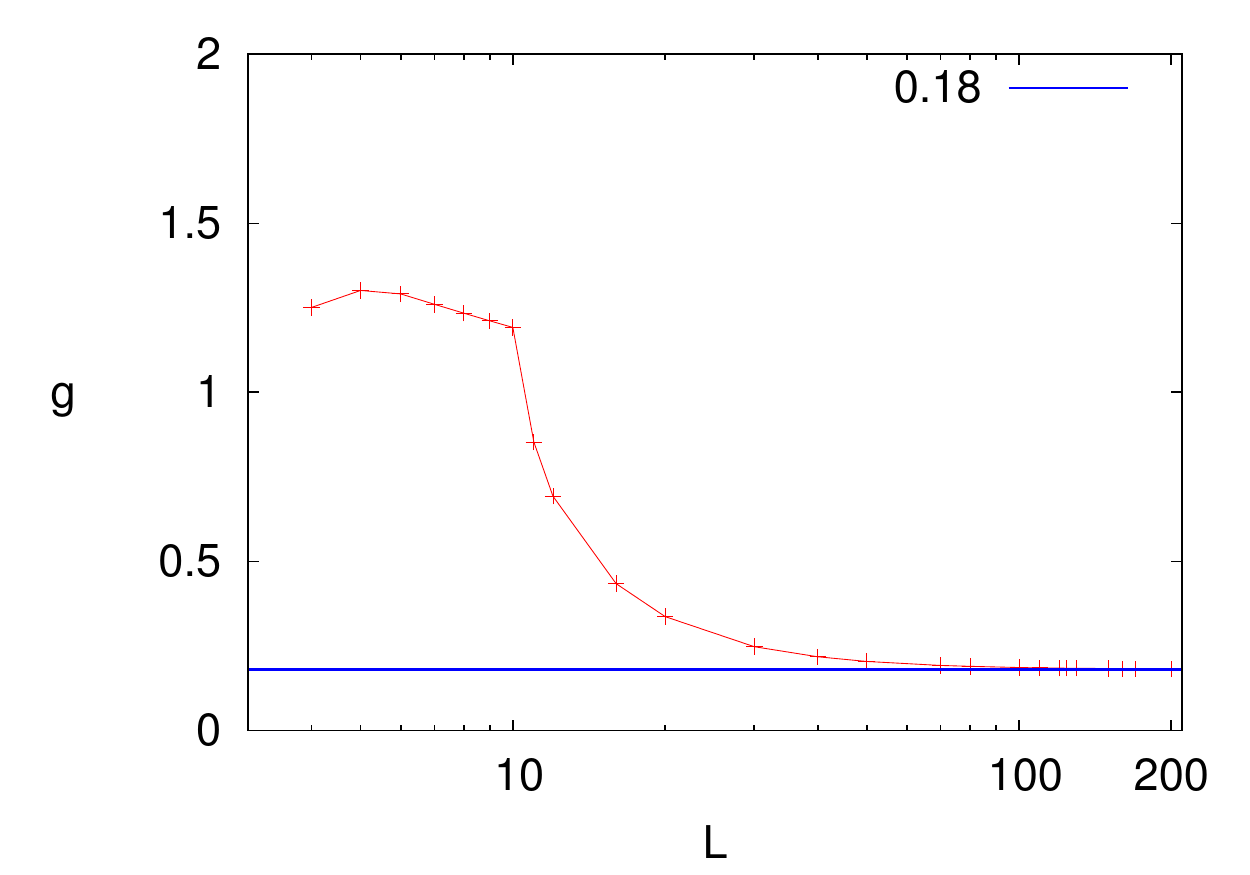}
\caption{(Color online) The gap $g$ in the $1-$particle sector of the XX chain with an incoherent one-way hopping (\ref{eq:oneway}). The kink at $L=10$ is because the eigenvalue responsible for $g$ goes from being complex to real.}
\label{fig:spsm}
\end{figure}

We have seen in the XX and XXZ models that the gap can be constant for a sufficiently weak bulk dissipation. Problem with these two cases is that the critical dissipation goes to zero in the thermodynamic limit. A natural question is: can one have a constant gap also for non-small dissipation? The answer is yes and we are going to give a simple example. Known are examples with only dissipation and no Hamiltonian~\cite{Diehl10,Horstmann13}. An example we are going to present has a nonzero Hamiltonian and nonzero dissipation. It is a XX chain with an incoherent ``hopping'' given by Lindblad operator
\begin{equation}
L_j=\sigma^+_j \sigma^-_{j+1},
\label{eq:oneway}
\end{equation}
at each site $j$, $\sigma^\pm_j=(\sx_j \pm {\rm i} \sy_j)/2$. The Liouvillian $\cL$ again conserves magnetization difference $z$ and particle number $r$. Numerically diagonalizing $\cL$ in a $r=1$ particle sector ($z=0$) one gets gaps shown in Fig.~\ref{fig:spsm} .
We can see that asymptotically the gap is independent of $L$. Because in the 1-particle sector interaction asymptotically does not matter, the same asymptotic gap would be obtained also in the XXZ chain with an incoherent hopping. Beware that if we would take the XX chain with an incoherent hopping in both directions (i.e., adding additional Lindblad operators $\sigma^-_j \sigma^+_{j+1}$) the gap in the 1-particle sector would asymptotically decay as $g \asymp 4\pi^2/L^2$, see also Ref.~\cite{eisler11}.

Another way of having a constant gap is to use a so-called thermal dissipators (also known as Davies generators~\cite{Davies74}) that can be derived in the limit of weak coupling to thermal reservoirs. The stationary state of such a master equation is thermal, though with Lindblad operators that are not local. For such models one can in certain cases rigorously prove that the gap is constant~\cite{Temme13}. 

\section{Boundary dissipation}

In this section we are going to study the gap in open systems with boundary dissipation, that is, with $\cLd$ acting in the thermodynamic limit nontrivially only on a finite number of sites. In all our cases $H$ will always be a spin$-1/2$ chain and dissipation will act on the leftmost and rightmost lattice sites. As mentioned, for boundary-driven open systems the gaps observed in the literature are all $\sim 1/L^3$ or smaller. Examples are $\sim 1/L^3$ in the XY chain with boundary magnetization driving~\cite{Prosen08,Bianchi14,Medvedyeva14} (or $\sim 1/L^5$ at nonequilibrium phase transition points~\cite{Pizorn08}; or even $\sim 1/L^7$ on the so-called resonances~\cite{Bojan}); $\sim 1/L^3$ is the scaling also for the magnetization-driven XXX model~\cite{JSTAT11}, or for the XXZ model with an incoherent hopping as a driving~\cite{Buca12}.

Considering these results one can ask whether the gap in a boundary-driven open system is perhaps always $\sim 1/L^3$ or smaller? As we will see the answer is no. But let us first present an argument that the gap can not be larger than $\sim 1/L$.

\subsection{Gap upper bound}

Let us have a one-dimensional system described by a local Liouvillian, $\cL=\sum_{j=1}^L \cL_j$, which is a sum of local term $\cL_j$, each of which acts nontrivially only on a fixed number of consecutive sites around site $j$ (Hilbert space dimension of a single site is finite, implying that $\cL_j$ are bounded, and the steady state is assumed to be unique). In addition, let all but a fixed number of $\cL_j$ be purely unitary, i.e., dissipation is present only in a fixed number of $\cL_j$. In the thermodynamic limit the number of consecutive sites without any dissipation therefore grows linearly with the system length $L$. The question we want to ask is: what is the fastest possible relaxation time, i.e., the largest possible gap, in such an open system?

We will show that the gap can not be larger than $\sim 1/L$, or, in other words, relaxation can not happen in a time that grows with $L$ slower than linearly. The argument is actually very simple. Because there are sites that are of distance $\sim L$ away from the nearest site with dissipation, a ``disturbance'' at that site can not dissipate in a time smaller than $\propto L$, i.e., the time a disturbance needs to get to a site with dissipation. We can also use a transport argument: for unitary evolution local conservation of energy holds and therefore, because the local energy current is a bounded operator, it will take at least a time $\propto L$ for the energy of the initial state to be dissipated, if we choose an initial state having total energy proportional to $L$. One could also rigorously formulate the above argument by e.g. using the Lieb-Robinson bound for open systems~\cite{Poulin10,Nacht11,Barthel12}. The Lieb-Robinson bound essentially formalizes a statement that there is a finite propagation speed in bounded locally-coupled systems. One consequence is that connected correlations of local operators get exponentially suppressed outside of a light cone, or, that the Heisenberg picture $A(t)$ of the initial local operator can be approximated by a part of $\cL$ with support inside the light cone (outside the light cone one has $A(t) \approx \1$). Taking a local $A(0) \perp \1$ with support on sites that are a distance $\sim L$ away from dissipation, one immediately sees that, because one has $A(t \to \infty)=0$, relaxation time can not be smaller than $\sim L$.

In next two subsections we are going to study by now familiar XX and XXZ models, showing that the Lieb-Robinson bound $g \sim 1/L$ is never reached, and then in the last subsection show some examples for which one does get $g \sim 1/L$.

\subsection{XX with boundary dephasing}

We take the XX chain (\ref{eq:XX}) with dephasing on the first and the last sites, that is, with only two Lindblad operators $L_1$ and $L_L$ from Eq.~(\ref{eq:dephasing}). Similarly as in the XX chain with dephasing on all sites, the gap is again the same in all $r$-particle sectors and we can limit our discussion to the 1-particle sector.

\begin{figure}[ht!]
\centerline{\includegraphics[width=1.65in]{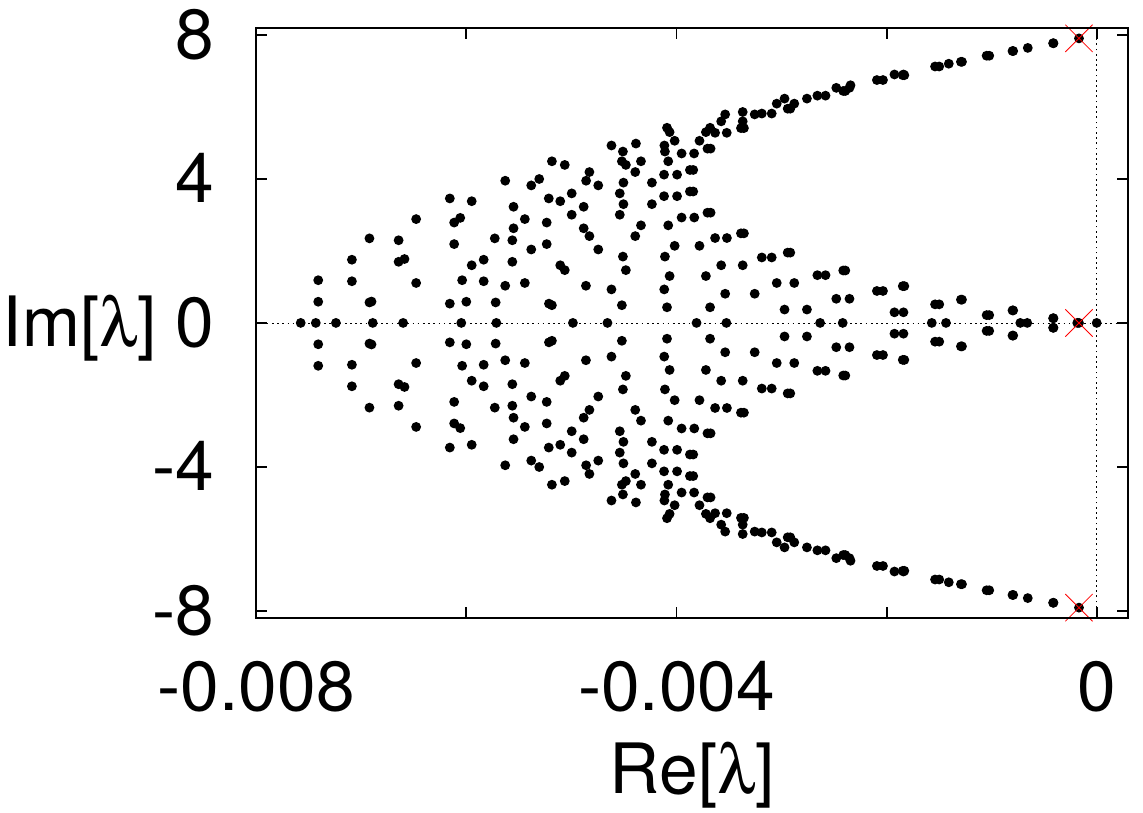}
\includegraphics[width=1.55in]{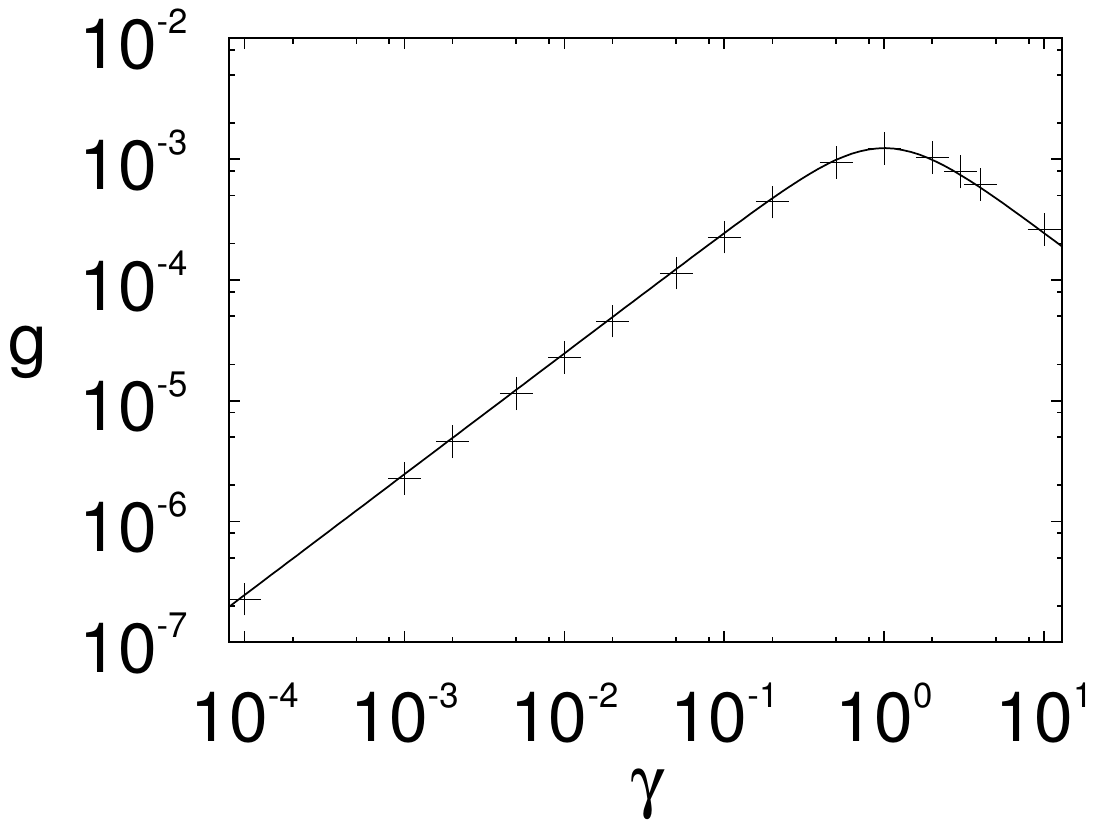}}
\caption{Left: Eigenvalues of $\cL$ in the 1-particle sector for the XX chain with boundary dephasing of strength $\gamma=0.01$. Three crosses are eigenvalues that have the largest real part and give $g$ (\ref{eq:boundaryXXsmallg}). Right: Numerical $g(\gamma)$ (symbols) for $L=40$ and prediction by Eq.(\ref{eq:boundaryXX}) (full curve).}
\label{fig:eigBoundXX}
\end{figure}
We are first going to use perturbation theory for small $\gamma$. An example of a spectrum of $\cL$ in the 1-particle sector is in Fig.~\ref{fig:eigBoundXX} . One observes that for small $\gamma$ the largest eigenvalue, determining the gap, is complex and has the largest absolute value of the imaginary part (two crosses in left Fig.~\ref{fig:eigBoundXX}). This pair of complex eigenvalues determines the gap all the way upto the maximum of $g(\gamma)$ (which for $L=40$ shown in right frame of Fig.~\ref{fig:eigBoundXX} happens around $\gamma \approx 1$). For larger $\gamma$ the gap is given by a real eigenvalue (the cross on a real line in Fig.~\ref{fig:eigBoundXX}) which though has in the leading order the same real part as the complex pair. This means that one can use simple nondegenerate perturbation theory to get the gap for small $\gamma$. The eigenvalue with the largest (or smallest) imaginary part of $\cLH$ is $\rho_0=\ket{\psi_1}\bra{\psi_L}$, where we denote eigenvalues of $H$ by $E_k=4\cos{q_k}, k=1,\ldots,L$, where $q_k:=\frac{\pi k}{L+1}$, and eigenstates by $\ket{\psi_k}=\sum_j c_{jk}\ket{j}, \quad c_r=\sqrt{2/(L+1)}\sin{q_{r}}$, where $\ket{j}$ denotes a state with all spins down apart from the $j$-th spin. The eigenvalue correction will then be given by $\kappa:=\tr{\rho_0 \cLd(\rho_0)}$, where dephasing $\cLd$ is a sum of nontrivial parts acting on the first and the last sites. $\cLd$ gives a non-zero value (equal to $-2\gamma$) only when acting on a non-diagonal operator on the first/last site, and therefore $\cLd(\rho_0)$ is a sum of terms that are either $\ket{j}\bra{1}$ or $\ket{j}\bra{L}$ (or their Hermitian adjoint), i.e., $\propto -2\gamma \sum_{j\neq 1} c_1 c_j^* \ket{1}\bra{j}+\cdots$, at the end resulting in
\begin{eqnarray}
\kappa &=&-\frac{8\gamma}{(L+1)^2}\Big[ 
-(2-\cos{q_2}-\cos{q_{2L^2}}) \sin^2{q_L}+  \nonumber \\ 
&& +\sum_{k=1}^L{s_{k,L}+s_{k,1}+s_{1,k}+s_{L,k}}\Big],
\label{eq:sum}
\end{eqnarray}
where $s_{k,r}:=\sin^2{q_k}\sin^2{q_{r L}}$. For large $L$ the term in the first line of Eq.~(\ref{eq:sum}) is subdominant, while the second line gives a large-$L$ expression for $\kappa$ and therefore also the gap, resulting in 
\begin{equation}
g=\gamma \frac{16\pi^2}{(L+1)^3}+{\cal O}(\gamma^2).
\label{eq:boundaryXXsmallg}
\end{equation}
It is instructive to qualitatively understand how the $\sim 1/L^3$ scaling comes about: we see that $s_{k,r}$ is just a product of eigenstate expansion coefficients $c_j$ and that the sum scales as $\sum_{k} |c_1|^2 |c_{kL}|^2 \sim |c_1|^2$, which in turn is the smallest overlap probability at the first site, giving one $1/L$ due to normalization and an additional $q_1^2 \sim 1/L^2$ due to the longest-wavelength eigenstate which is the slowest to relax. Such a scenario has already been observed~\cite{Medvedyeva14} in the XX chain with boundary injection of particles, i.e., a model with Lindblad operators $\sim \sigma^\pm$ at the boundaries instead of our $\sim \sz$, making it exactly solvable. One can in fact argue that such $\sim 1/L^3$ scaling is generic for boundary driven models with integrable Hamiltonian $H_0$ having a plane-wave like longest-wavelength eigenstates.

For large $\gamma$ one on the other hand expects the gap to decay with dephasing strength as $\sim 1/\gamma$ because of an effective decoupling of the system and as a consequence increasing relaxation time. One can extend small-$\gamma$ expression (\ref{eq:boundaryXXsmallg}) to also have the correct large-$\gamma$ behavior by writing,
\begin{equation}
g=\frac{\gamma}{1+\gamma^2}\frac{16\pi^2}{L^3}.
\label{eq:boundaryXX}
\end{equation}
One can see in right Fig.~\ref{fig:eigBoundXX} that this expression indeed fits well numerical results (small discrepancy seen is due to a subleading $\sim 1/L^4$ correction to Eq.~(\ref{eq:boundaryXX})). Observe that for the boundary driven XX system, as opposed to the XX with bulk dephasing, there is no transition in the scaling of the gap as one varies $\gamma$.

\subsection{XXZ with boundary dephasing}

For the XXZ model with boundary dephasing we take a standard XXZ Hamiltonian (\ref{eq:XXZ}) and the same two Lindblad operators $L_1$ and $L_L$ (\ref{eq:dephasing}). We begin by discussing weak dephasing.

For weak dephasing one observes that the gap comes from real eigenvalue of $\cL$ and one therefore has to use degenerate perturbation theory in an appropriate $r$-particle subspace (again, always limiting to $z=0$). First order already gives a non-zero contribution and so the procedure is very similar to the one already used for the XXZ chain with bulk dephasing, leading to Eq.~(\ref{eq:R}). Repeating the same steps now for $\cLd$ that acts only on the boundary two sites, one gets
\begin{equation}
R_{j,k}=|\bracket{\psi_j}{\sz_1}{\psi_k}|^2+|\bracket{\psi_j}{\sz_L}{\psi_k}|^2-2 \delta_{j,k},
\label{eq:boundaryR}
\end{equation} 
where $\ket{\psi_k}$ are eigenstates of the XXZ chain in an appropriate $r$-particle sector. Eigenvalues of $R$ are for small $\gamma$ equal to the largest real eigenvalues (divided by $\gamma$) of $\cL$. Doing the calculation in the 1-particle sector one gets
\begin{equation} 
g=\frac{16\pi^2}{(1+L)^3}\frac{\gamma}{(1+\Delta)^2}+{\cal O}(\gamma^2).
\label{eq:boundXXZweak1mag}
\end{equation}
\begin{figure}[t!]
\centerline{\includegraphics[width=1.65in]{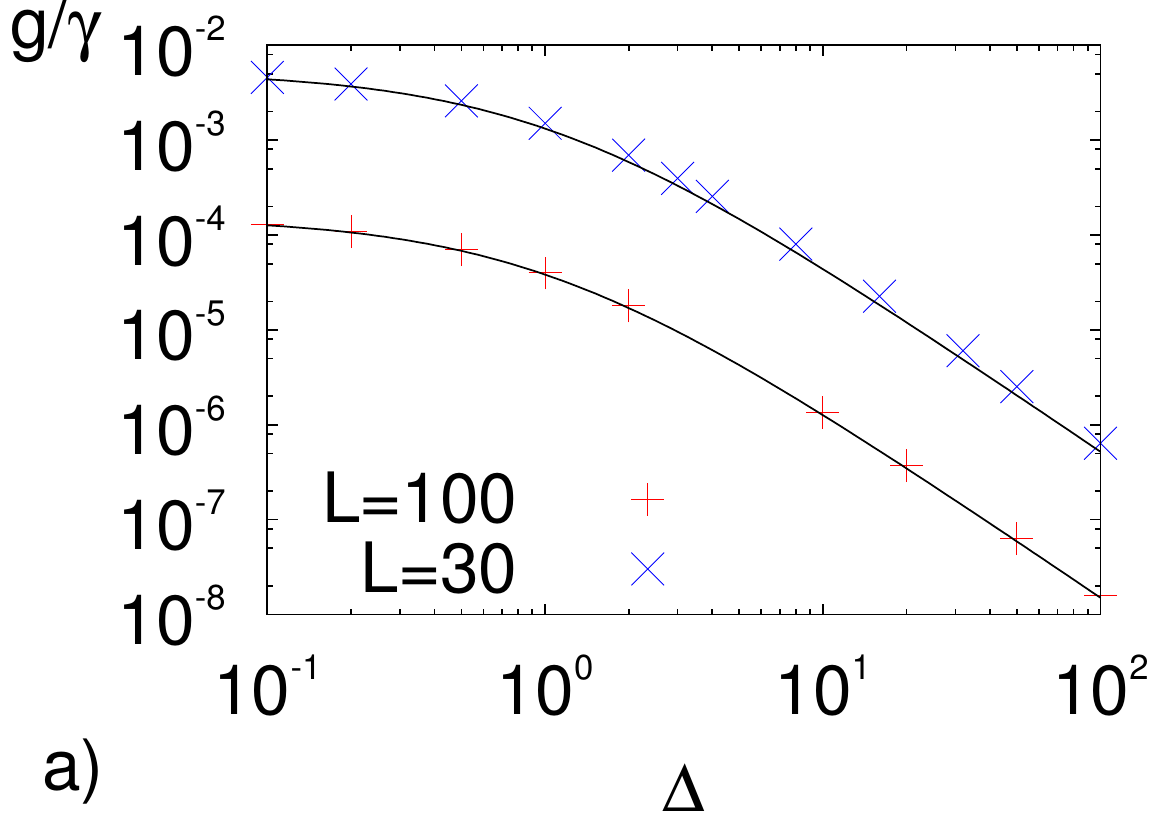}
\includegraphics[width=1.55in]{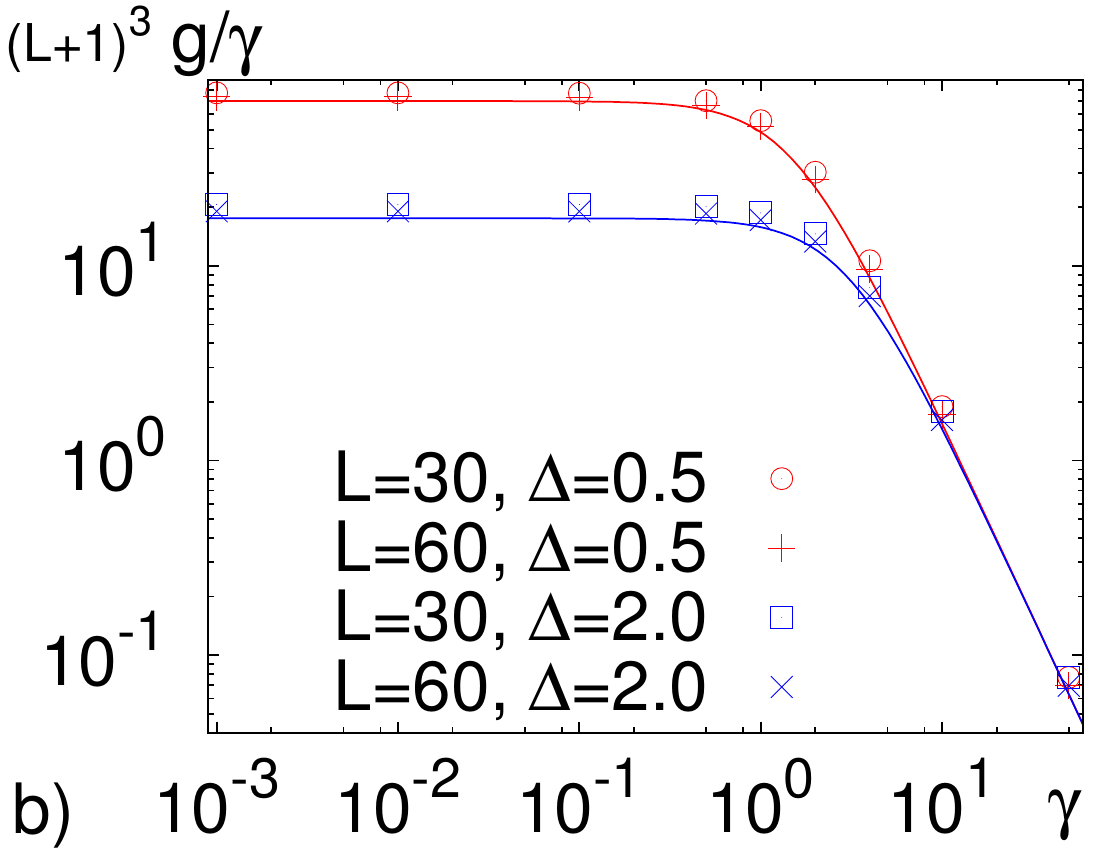}}
\caption{(Color online) 1-particle sector of the XXZ chain with boundary dephasing. (a) Numerically calculated gap $g$ for small $\gamma=0.01$ and the theory (full curves, Eq.~(\ref{eq:boundXXZweak1mag})). (b) $g(\gamma)$ always scales as $\sim 1/L^3$; symbols are numerical values for two different sizes and $\Delta$, full curves are Eq.~(\ref{eq:boundXXZ1mag}).}
\label{fig:boundXXZ}
\end{figure}
For small $\gamma$ and in the 1-particle sector the gap therefore scales as $\sim 1/L^3$ irrespective of the value of $\Delta$. For a non-small $\gamma$ expression that fits numerics well and has a correct weak-dephasing limit (\ref{eq:boundXXZweak1mag}) is
\begin{equation}
g\approx \frac{16\pi^2}{(L+1)^3}\frac{\gamma}{\gamma^2+(1+\Delta)^2}.
\label{eq:boundXXZ1mag}
\end{equation}
We see in Fig.~\ref{fig:boundXXZ}b that the perturbative result in Eq.~(\ref{eq:boundXXZweak1mag}), i.e., horizontal dependence for small $\gamma$ in the figure, holds up-to an $L$-independent $\gamma$.

\begin{figure}[t!]
\centerline{\includegraphics[width=1.6in]{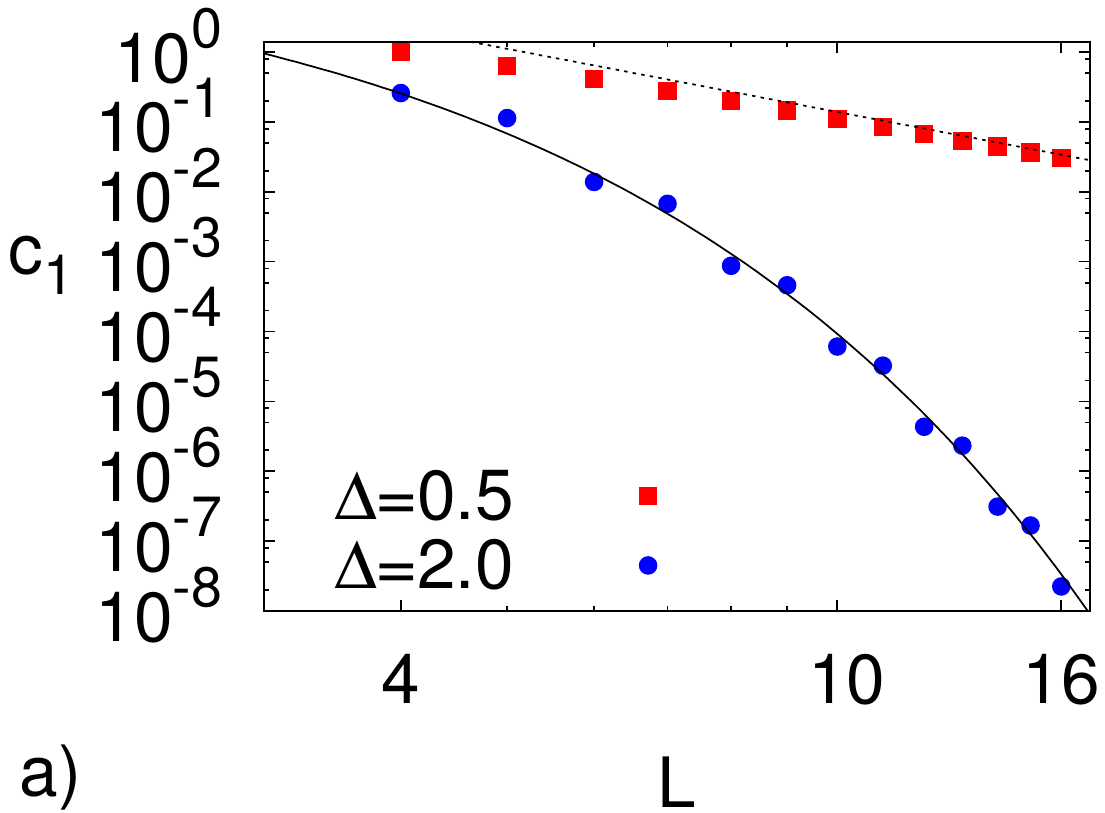}
\includegraphics[width=1.6in]{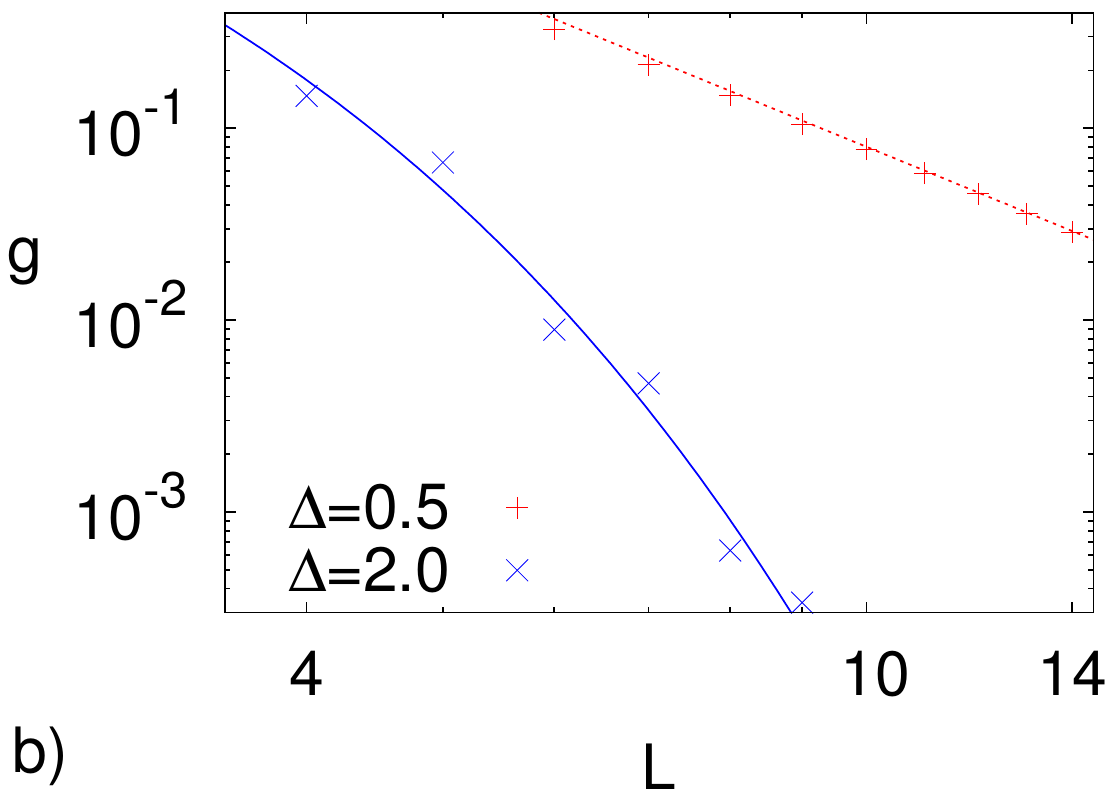}}
\caption{(Color online) Half-filling sector of the XXZ chain with boundary dephasing. (a) The largest non-zero eigenvalue of $R$, Eq.~(\ref{eq:boundaryR}), determining the gap for small $\gamma$ (\ref{eq:XXZ-weak}). Asymptotic behavior is denoted by dotted line ($=139/L^3$) and full curve ($\sim \exp{(-1.32L)}$). (b) The gap for non-small $\gamma=1$. Asymptotically the scaling is the same as for small $\gamma$, namely, the dotted line is $80/L^3$ ($\Delta=0.5$) and the full curve is $\sim \exp{(-1.32L)}$ ($\Delta=2.0$).}
\label{fig:boundXXZhalf}
\end{figure}

We now move to the half-filling sector. In Fig.~\ref{fig:boundXXZhalf} we show the results. For large $L$ the gap scales as $g \sim 1/L^3$ in the gapless phase $\Delta<1$, while it is exponentially small, $g \sim \exp{(-\alpha L)}$, in the gapped phase of $\Delta>1$~\cite{foot4}. Comparing gaps in the 1-particle and the half-filling sector one also observes (data not shown) that for $\Delta<1$ the gap in the 1-particle sector is the smallest of the two, while for $\Delta>1$ the smallest gap is from the half-filling sector. For arbitrary $\gamma$ the global gap therefore scales as $\sim 1/L^3$ for $\Delta<1$ and $\sim \exp{(-\alpha L)}$ for $\Delta>1$, see also Table~\ref{tab:bound} . 

An exponentially small gap can be most easily understood for small $\gamma$ via matrix $R$. Without loss of generality we limit to even $L$ and focus on the half-filling sector. For matrix elements of $R$ eigenstates of $H$ and matrix elements of Lindblad operators is what matters. In the spectrum of $H$ there are two almost degenerate eigenstates $\psi_{1,2}$ which are the most important for the gap, i.e., for the largest nontrivial eigenvalue of $R$. For large $\Delta$ they are a symmetric and antisymmetric combination of a domain wall in the middle of the chain, $\ket{\psi_1}\sim \ket{R}+\ket{L}$ and $\ket{\psi_2}\sim \ket{R}-\ket{L}$, where we denoted $\ket{L}\equiv \ket{11\ldots1 00\ldots 0}$ and $\ket{R} \equiv \ket{00\ldots0 11\ldots 1}$ ($\ket{L}$ is a state with the left-half of spins down, and $\ket{R}$ a state with the right-most half of spins down). For finite $\Delta$ domain wall states $\ket{R,L}$ differ from a perfect wall only within a localization length $\sim 1/\ln{\Delta}$ of the middle spin, see e.g. the Appendix in Ref.~\cite{NDC09}. For $\Delta>1$ the two states $\psi_{1,2}$ are therefore ``localized'' around the site $n/2$, their energy is almost degenerate, they are the highest energy states (in the half-filling sector), and they are gapped by $\approx 2\Delta$ from the rest of the spectrum. They have an important property that $\sz_1 \ket{\psi_1} \approx \ket{\psi_2}$ and $\sz_1 \ket{\psi_2} \approx \ket{\psi_1}$, i.e., a subspace spanned by $\psi_{1,2}$ is invariant under $\sz_1$. As a consequence, overlaps $\bracket{\psi_{k\neq 1,2}}{\sz_1}{\psi_{1,2}}$ are exponentially small in $L$ and in the matrix $R$ a $2\times 2$ block corresponding to these two eigenmodes is decoupled from the rest. This $2\times 2$ block is of the form $-2\1+(2-\epsilon)\sx$ resulting in two eigenvalues $-4+\epsilon$ and $-\epsilon$, with the corresponding Liouvillian decay eigenmodes $\ket{\psi_1}\bra{\psi_1}+\ket{\psi_2}\bra{\psi_2}$ for eigenvalue $-\epsilon$, and $\ket{\psi_1}\bra{\psi_1}-\ket{\psi_2}\bra{\psi_2}$ for a symmetric partner at $-4+\epsilon$ (remember that the steady state is a uniform mixture of all projectors, $\sum_k \ket{\psi_k}\bra{\psi_k}$). The gap is thus equal to $\epsilon$ which is in turn determined by the localization length of two localized eigenmodes. The rate $\alpha$ in $g \sim \exp{(-\alpha L)}$ is therefore proportional to the inverse localization length, resulting in $\alpha \propto \ln{\Delta}$. Exponentially slow relaxation found for $\Delta>1$ could be interesting for instance for quantum memory -- we see that the domain wall can support a one-qubit quantum memory formed out of $\psi_{1,2}$ which is exponentially resilient to dephasing. Unfortunately, as we shall see in the next magnetization-driven model, this resilience is lost for boundary dissipation in a transverse direction. We note that one can also get exponentially small gap because of localization due to disorder, an example being a boundary driven XY chain with a disordered magnetic field~\cite{Prosen08}.

\subsection{Magnetization-driven XXZ}

So-far we have used dephasing as our canonical example of dissipation because it eased up theoretical analysis. Steady states are in those cases rather simple, namely uniform mixtures of all projectors in a respective symmetry subspace (Appendix~\ref{app:A}). Much more interesting steady states arise if one has boundary driving that breaks symmetries, for instance imposing a left-right asymmetry by having different driving at chain ends. In such a case the steady state will be a genuinely nonequilibrium state with nonzero currents flowing through the system.

In this subsection we are going to study one such example, on one hand to contrast the obtained scaling with that of a boundary dephasing case and on the other hand to calculate the gap for a physically much relevant open system. We shall take the same XXZ model as in previous sections (\ref{eq:XXZ}), without dephasing, but instead with a boundary magnetization driving described by 2 Lindblad operators at each end,
\begin{eqnarray}
L_1=\sqrt{\Gamma(1+\mu)}\,\sigma^+_1,\quad L_2=\sqrt{\Gamma(1-\mu)}\,\sigma^-_1,\nonumber \\
L_3=\sqrt{\Gamma(1-\mu)}\,\sigma^+_L,\quad L_4=\sqrt{\Gamma(1+\mu)}\,\sigma^-_L.
\label{eq:mu}
\end{eqnarray}
Driving parameter $\mu$ parametrizes the asymmetry between injection and absorption of particles at the boundary, trying to impose expectation value of $\sz$ equal to $\pm \mu$ at chain ends, while $\Gamma$ is the coupling strength. Nonzero $\mu$ therefore causes a nonzero magnetization gradient along the chain. We are going to use $\mu=0.1$, however, the scalings observed (e.g., Table~\ref{tab:bound}) are independent of the precise value of $\mu$ as long as $\mu$ is not too close to $\mu=1$ for which one gets a blockade of transport due a step-like magnetization profile resulting also in an exponentially small gap leading to exponentially slow relaxation~\cite{NDC}. The Liouvillian of such magnetization-driven XXZ model still conserves $z$, but not anymore $r$ (\ref{eq:const}). Note that $\Gamma$ plays the role of dissipation strength, similarly to $\gamma$ in the case of dephasing dissipation -- small $\Gamma$ means weak dissipation, e.g., weak external coupling. Dissipation strength in the paper is therefore determined either by $\gamma$ for dephasing (\ref{eq:dephasing}), or by $\Gamma$ for ``magnetization'' driving in Eq.~(\ref{eq:mu}).

\begin{figure}[t!]
\includegraphics[width=3.1in]{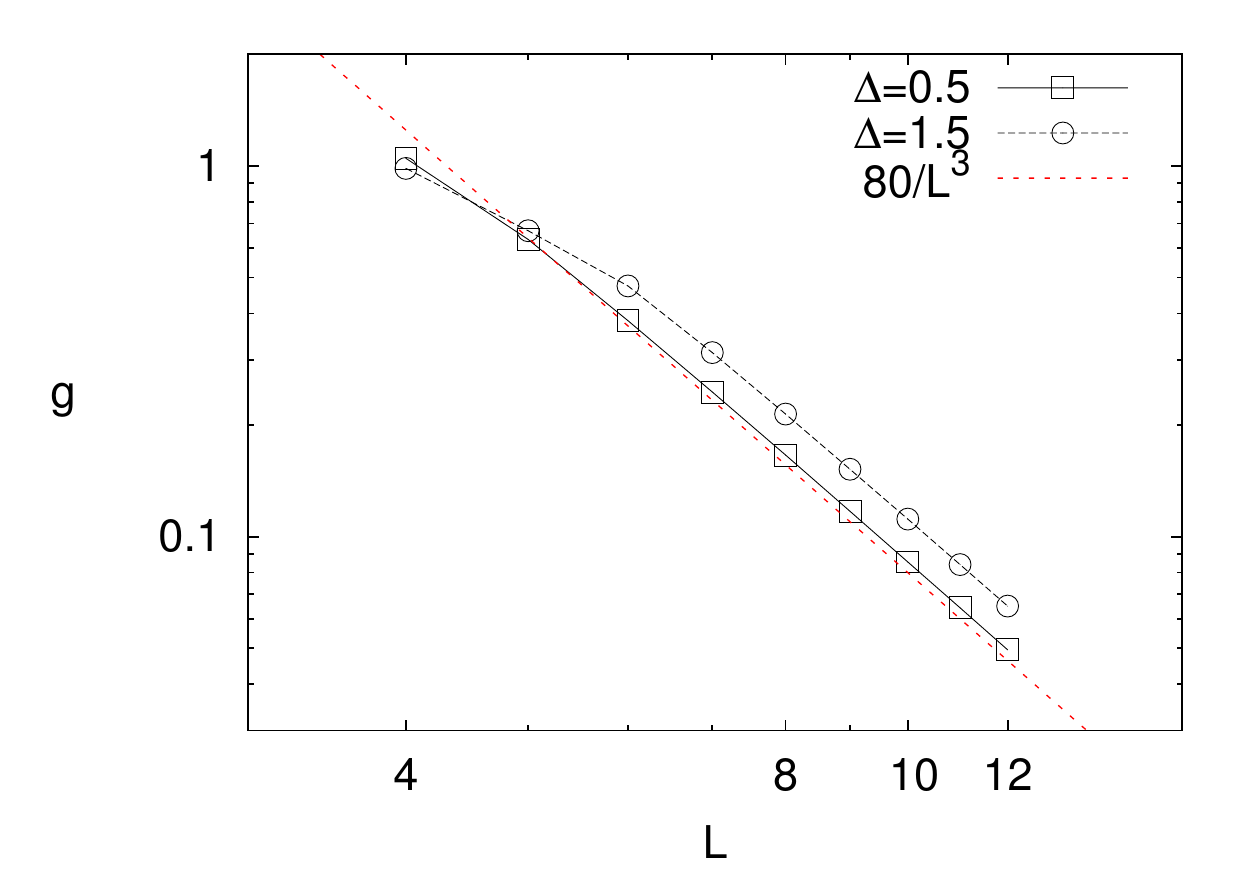}
\caption{The gap $g$ for the XXZ model and boundary driving (\ref{eq:mu}) with $\mu=0.1$ and $\Gamma=1$. Straight dashed line suggests asymptotic scaling $g \propto 1/L^3$.}
\label{fig:drivXXZ}
\end{figure}
Numerically calculated gaps are shown in Fig.~\ref{fig:drivXXZ} . The gap $g$ looks to have a nice $\sim 1/L^3$ scaling regardless of the anisotropy $\Delta$. We can compare the gap obtained here to the one obtained in previous subsection for the XXZ model driven with boundary dephasing (Fig.~\ref{fig:boundXXZhalf}): in the gapless phase of $\Delta <1$ the scaling is in both cases $\sim 1/L^3$, while in the gapped phase of $\Delta >1$ the gap is here $\sim 1/L^3$ while it was exponentially small for the boundary dephasing case. We see that the scaling of the gap can change already by changing boundary terms only.

Analyzing the gap for small coupling $\Gamma$, i.e., small dissipation, again requires application of a degenerate perturbation theory. Using Lindblad operators (\ref{eq:mu}) the perturbation matrix $R$ is in this case
\begin{eqnarray}
R_{jk}&=&2(1+\mu)|\bracket{\psi_j}{\sigma_1^+}{\psi_k}|^2+2(1-\mu)|\bracket{\psi_j}{\sigma_1^-}{\psi_k}|^2+\nonumber \\
&+&2(1-\mu)|\bracket{\psi_j}{\sigma_L^+}{\psi_k}|^2+2(1+\mu)|\bracket{\psi_j}{\sigma_L^-}{\psi_k}|^2-\nonumber \\
&-&2\mu \delta_{jk}\bracket{\psi_j}{\sz_1-\sz_L}{\psi_j}-4 \delta_{jk}.
\label{eq:muR}
\end{eqnarray}
The spectrum of such $R$ for the XXZ chain is in fact independent of $\mu$. For $\mu=0$ the above $R$ can be further simplified, taking into account also that $H$ is real, results in
\begin{equation}
R_{jk}=2|\bracket{\psi_j}{\sx_1}{\psi_k}|^2+2|\bracket{\psi_j}{\sx_L}{\psi_k}|^2-4\delta_{jk}.
\label{eq:muRsimpl}
\end{equation}
Such $R$ is real and symmetric with the eigenvector corresponding to eigenvalue $0$ being a uniform superposition of all basis states (we remind that perturbative $R$ has always one eigenvalue equal to $0$ with the corresponding eigenvector though being in general more complicated). The gap (i.e., $c_1$) for small coupling $\Gamma$ of magnetization driving therefore depends on fluctuations of $\sx$ at the boundary two sites; this can be contrasted with the dephasing case (\ref{eq:boundaryR}) where fluctuations of $\sz$ matter.
\begin{figure}[!h]
\centerline{\includegraphics[width=1.55in]{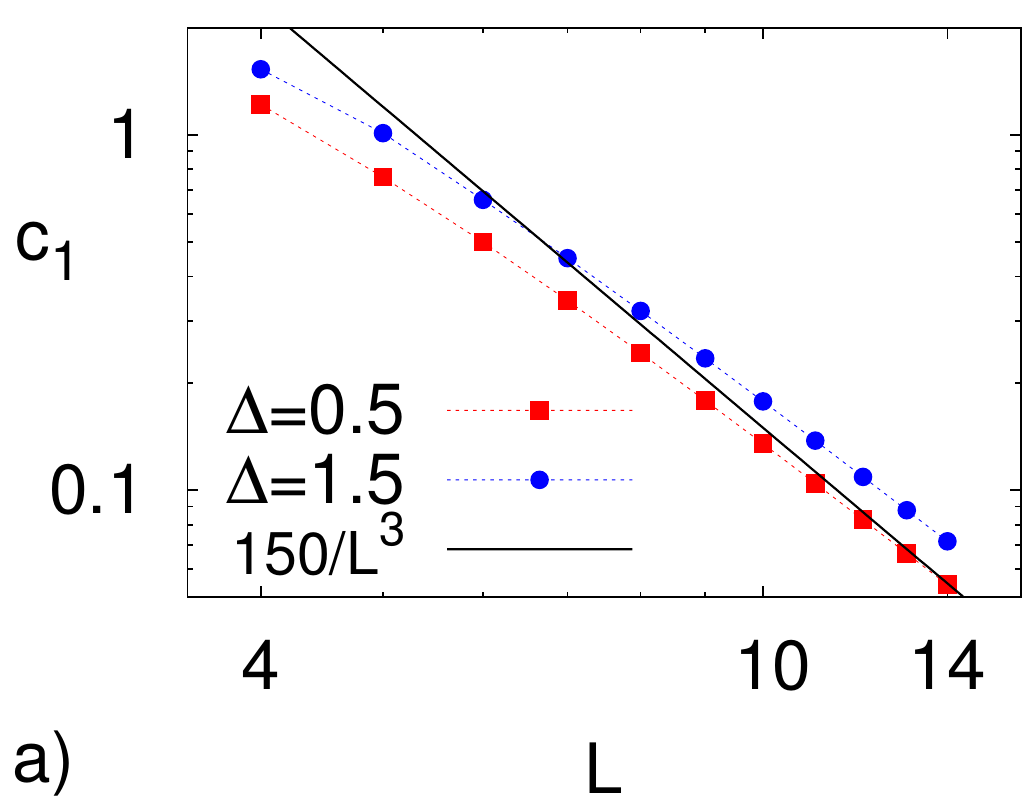}
\includegraphics[width=1.65in]{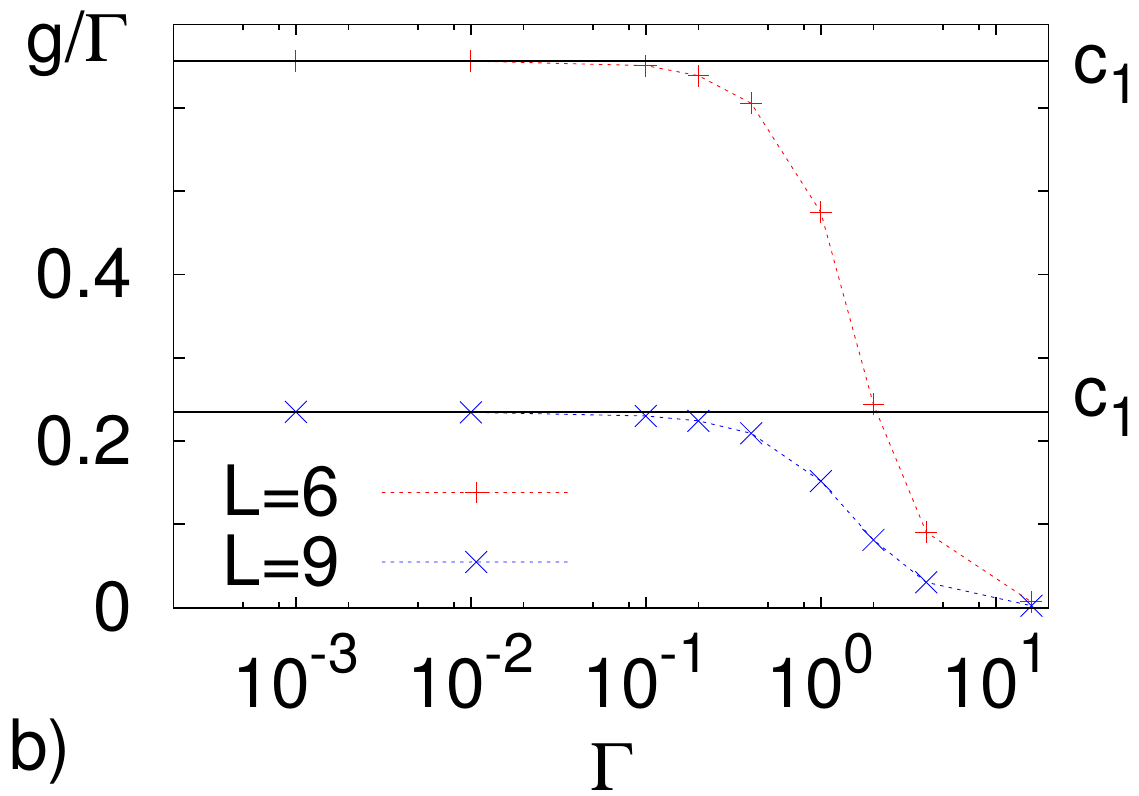}}
\caption{(Color online) The XXZ chain with magnetization driving and small dissipation $\Gamma$. (a) The largest non-zero eigenvalue $c_1$ of $R$ (Eq.~(\ref{eq:muR})) determining the gap for small $\Gamma$. Full line suggests asymptotic scaling $\sim 1/L^3$. (b) Dependence of exact gap on $\Gamma$ and perturbative $c_1$ from frame (a), all for $\Delta=1.5$ (for $\Delta=0.5$ behavior is similar). Deviations from $c_1$ begin at an $L$-independent $\Gamma$.}
\label{fig:weakXXZmu}
\end{figure}
In Fig.~\ref{fig:weakXXZmu} we show results for $c_1$ and verification of its validity. We can see that for small $\Gamma$ the gap also scales as $\sim 1/L^3$ irrespective of $\Delta$ and that approximation $g \approx \Gamma c_1$ holds upto an $L$-independent $\Gamma$. What is interesting is that, compared to the XXZ model with boundary dephasing we changed only boundary driving, that is, we modified only action of $\cL$ on $2$ out of $L$ sites, and nevertheless the scaling of gap for $\Delta>1$ changes from exponential to algebraic. On the level of perturbative matrix $R$ this can be understood as being due to the breaking of the underlying symmetry of two domain-wall states. Because the location of a symmetry-breaking perturbation is important (\ref{eq:muRsimpl}), having $\sigma^\pm$ dissipation at sites other than the boundary ones could still result in an exponentially small gap, see also recent Ref.~\cite{Carmele} for a study of stability of edge modes to Markovian dissipation.

We note that the scaling of the average magnetization current in the nonequilibrium steady state looks diffusive~\cite{PRL11} for this model in the gapped regime while higher current cumulants show anomalous non-diffusive scaling~\cite{PRB14}. Non-diffusive scaling $g \sim 1/L^3$ of the gap observed here (irrespective of $\Delta$) is perhaps an additional indication speaking in favor of anomalous transport properties of the gapped Heisenberg model.

\subsection{Fast boundary-driven relaxation}

So-far in all models studied the gap was $\sim 1/L^3$ or smaller. On the other hand the Lieb-Robinson argument puts a larger upper bound on the gap of $\sim 1/L$. One can wonder whether this upper bound can be saturated? We are going to demonstrate that there are models with $g \sim 1/L$. To achieve that we are going to take chaotic models with boundary driving.

\subsubsection{Staggered XXZ with boundary dephasing}

We are first going to consider the XXZ chain in a staggered magnetic field,
\begin{equation}
H=\sum_{j=1}^{L-1} \sx_j \sx_{j+1}+\sy_j \sy_{j+1}+\Delta \sz_j \sz_{j+1}+\sum_{j=1}^L b_j \sz_j,
\label{eq:XXZstagg}
\end{equation}
with staggered field having a period of 3 sites, $b_j=(-1,-\frac{1}{2},0,-1,-\frac{1}{2},0,\ldots)$, for which the Hamiltonian is quantum chaotic~\cite{PRE10} and shows diffusive magnetization transport~\cite{JSTAT09}. Dissipation will be dephasing on the 1st and the last site. Similarly as in other models, $z$ and $r$ are conserved and one can look at the gap in each $r$-particle sector with $z=0$.

\begin{figure}[t!]
\includegraphics[width=3.2in]{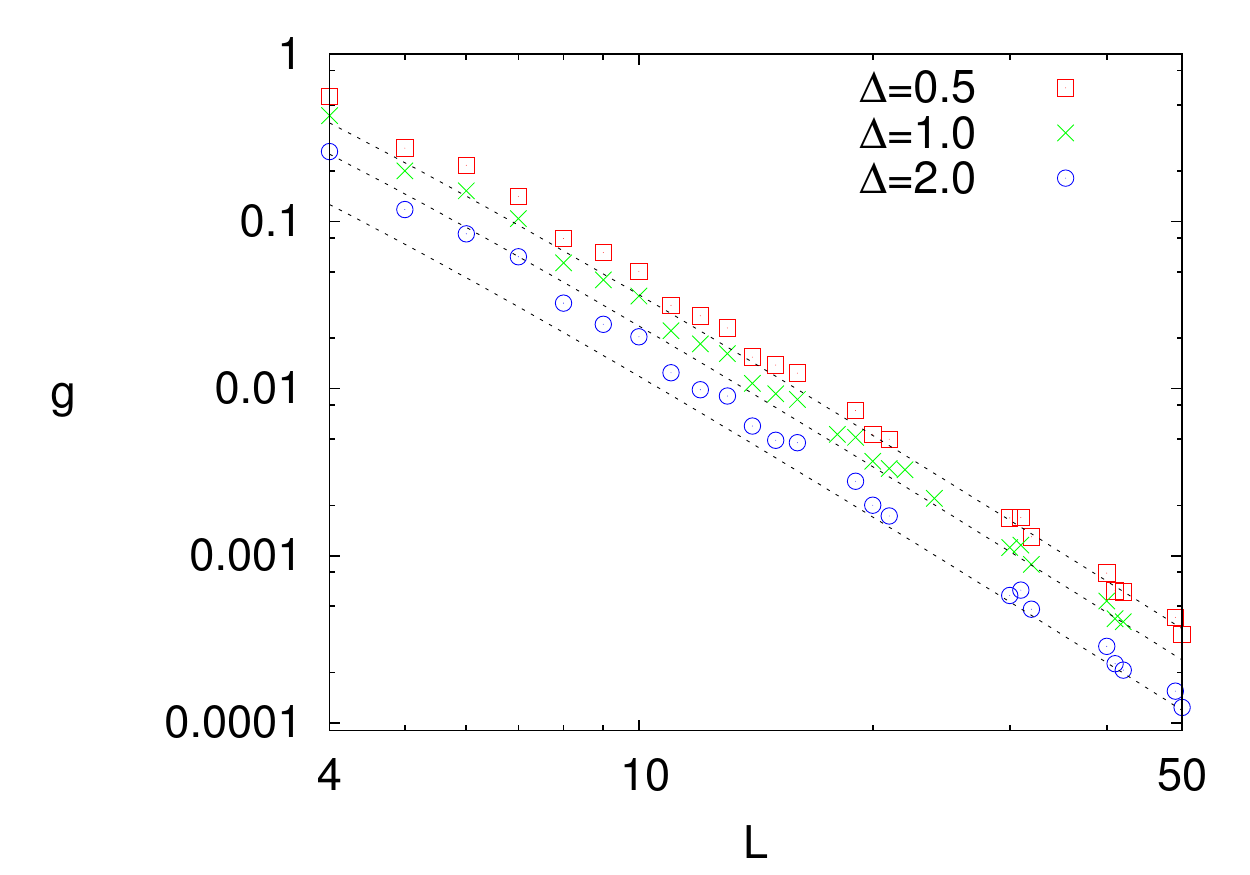}
\caption{The gap $g$ for the staggered XXZ chain with boundary dephasing ($\gamma=1$) in the 1-particle sector. Dashed black lines are Eq.(\ref{eq:boundXXZ1mag}), having asymptotic decay $\sim 1/L^3$.}
\label{fig:stagg1mag}
\end{figure}

In the 1-particle sector the staggered field has no influence on the asymptotic gap and Eq.(\ref{eq:boundXXZ1mag}), with scaling $\sim 1/L^3$ describes asymptotic $g$ well, see Fig.~\ref{fig:stagg1mag} .

\begin{figure}[!h]
\centerline{\includegraphics[width=1.55in]{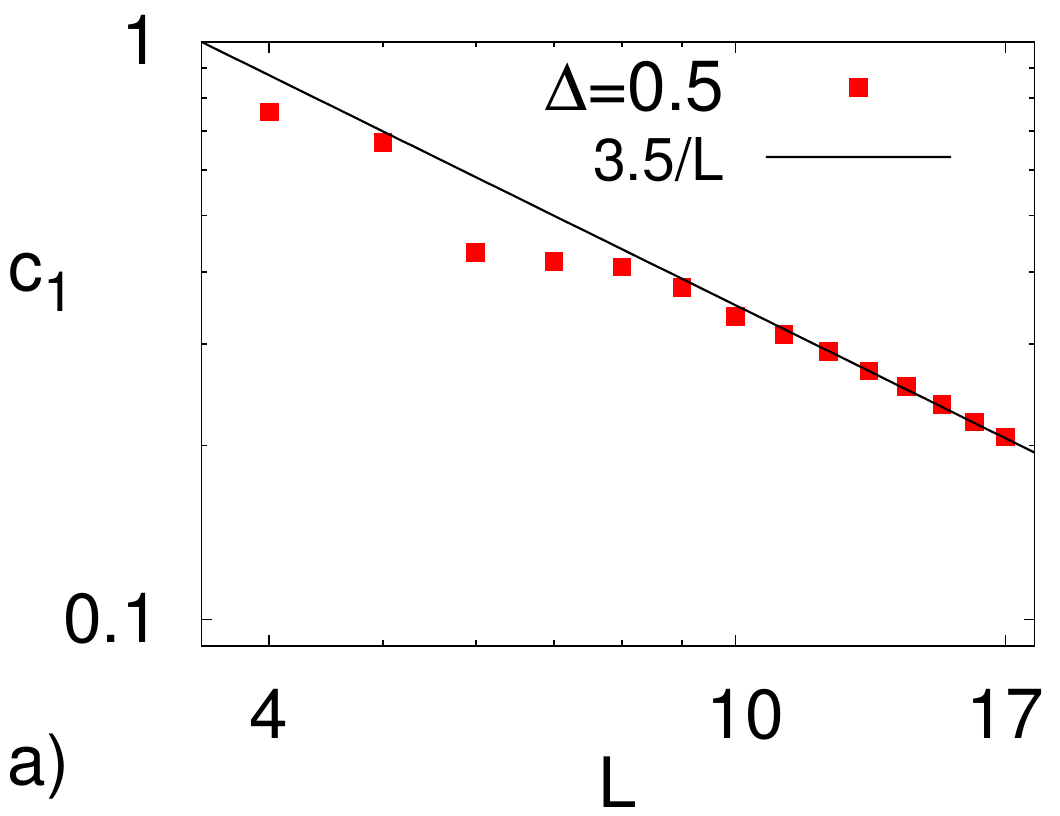}
\includegraphics[width=1.65in]{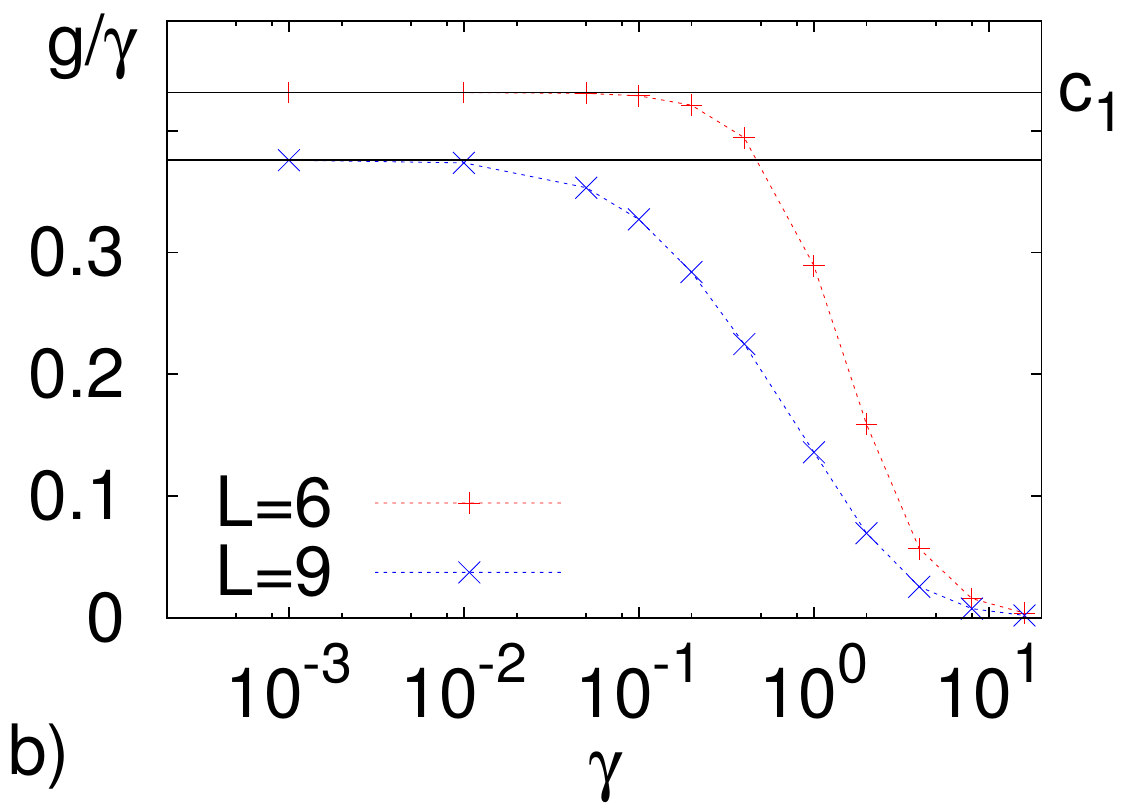}}
\caption{(Color online) Staggered XXZ chain with $\Delta=0.5$ and weak boundary dephasing. (a) The largest non-zero eigenvalue of $R$, Eq.~(\ref{eq:boundaryR}), determining the gap for small $\gamma$. Asymptotic behavior (full line) is $\approx 3.5/L$.  (b) With increasing size the convergence radius $\gamma_{\rm c}$ upto which perturbation theory holds decreases.}
\label{fig:staggXXZpert}
\end{figure}

In the half-filling sector we are first going to evaluate perturbation theory for small $\gamma$. The procedure is exactly the same as for the XXZ model without the field -- one has to calculate the largest non-zero eigenvalue $c_1$ of the matrix $R$ written in Eq.~(\ref{eq:boundaryR}) using eigenstates of the staggered XXZ chain -- with the gap then being given by $g \approx \gamma c_1$. Results are in Fig.~\ref{fig:staggXXZpert} . For $\Delta <1$ eigenvalue $c_1$ scales as $\sim 1/L$, while for $\Delta >1$ (data not shown) it is exponentially small in $L$. 
From Fig.~\ref{fig:staggXXZpert}b we also see that the validity of small-$\gamma$ approximation shrinks with growing $L$.

\begin{figure}[t!]
\includegraphics[width=3.2in]{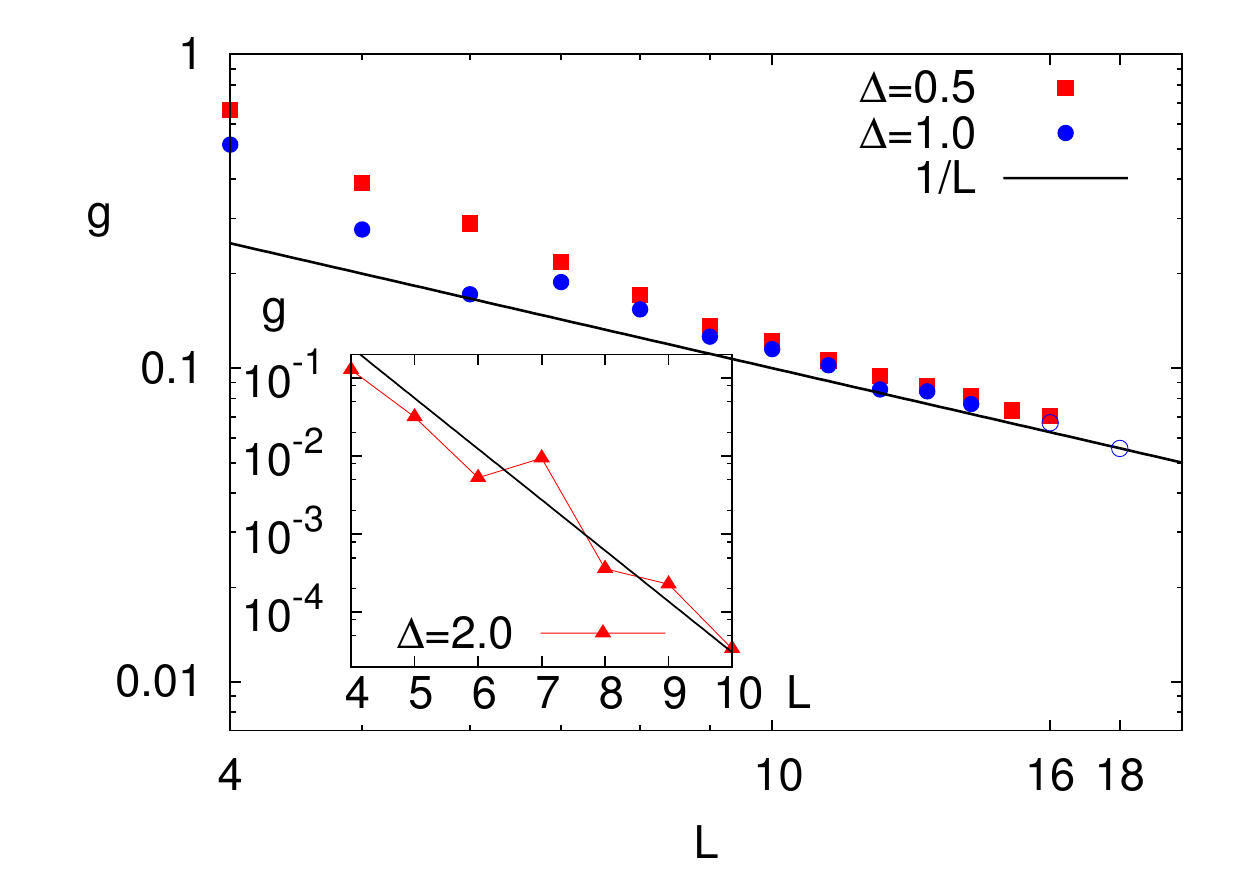}
\caption{(Color online) The gap $g$ for the staggered XXZ chain with boundary dephasing ($\gamma=1$) in the half-filled sector. The full black line in the main plot is $1/L$; in the inset it is $\propto \exp{(-1.5 L)}$. Full symbols are obtained by exact diagonalization, empty with tDMRG.}
\label{fig:stagghalf}
\end{figure}

For non-small $\gamma$ we numerically calculated gaps in the half-filling sector, see Fig.~\ref{fig:stagghalf} . Besides exact diagonalization we also used tDMRG to obtain the gap, observing relaxation of $\tr{\rho(t) \sx_{L/2} \sx_{L/2+1}}$, however, the required matrix dimension increases with $L$ rapidly and one can not go to large system sizes. We see that for $\Delta \le 1$ the gap scales as $\sim 1/L$, which is different than without the staggered field, when it is $\sim 1/L^3$ (Fig.~\ref{fig:boundXXZhalf}b). For $\Delta >1$ though the gap is exponentially small, the same as without the staggered field. Explanation is again in terms of localized modes.
In the gapless phase the smallest gap is from the 1-particle sector, in the gapped phase it is from the half-filling sector. The global gap therefore scales as $\sim 1/L^3$ for $\Delta \le 1$ and as $\sim \exp{(-\alpha L)}$ in the gapped phase. 

The fact that the gap remains $\sim 1/L^3$ in the 1-particle sector despite chaoticity is not surprising. What is interesting and puzzling is that in the half-filling sector and for $\Delta \le 1$ the gap looks $\sim 1/L$ (at least for the sizes available, Fig.~\ref{fig:stagghalf}), despite chaoticity and diffusive transport. We do not understand at present how such fast relaxation is compatible with diffusion, we note though that relaxation towards the steady state and transport properties of the steady state are in principle two separate properties.

\subsubsection{Magnetization-driven staggered XXZ}

It could be that the above fast relaxation is due to conservation of $r$. We therefore take the same XXZ chain with staggered field as above, Eq. (\ref{eq:XXZstagg}), but this time with boundary magnetization driving of Eq.~(\ref{eq:mu}) instead of with dephasing. The value of driving is chosen to be $\mu=0.1$. Now the Liouvillian conserves only $z$, and the reported gaps are for the $z=0$ sector (eigenvalues in other sectors have larger gaps; sectors $z \neq 0$ also do not contain any steady state). The steady state is nontrivial and represents a nonequilibrium state with nonzero magnetization current.
\begin{figure}[b!]
\centerline{\includegraphics[width=1.55in]{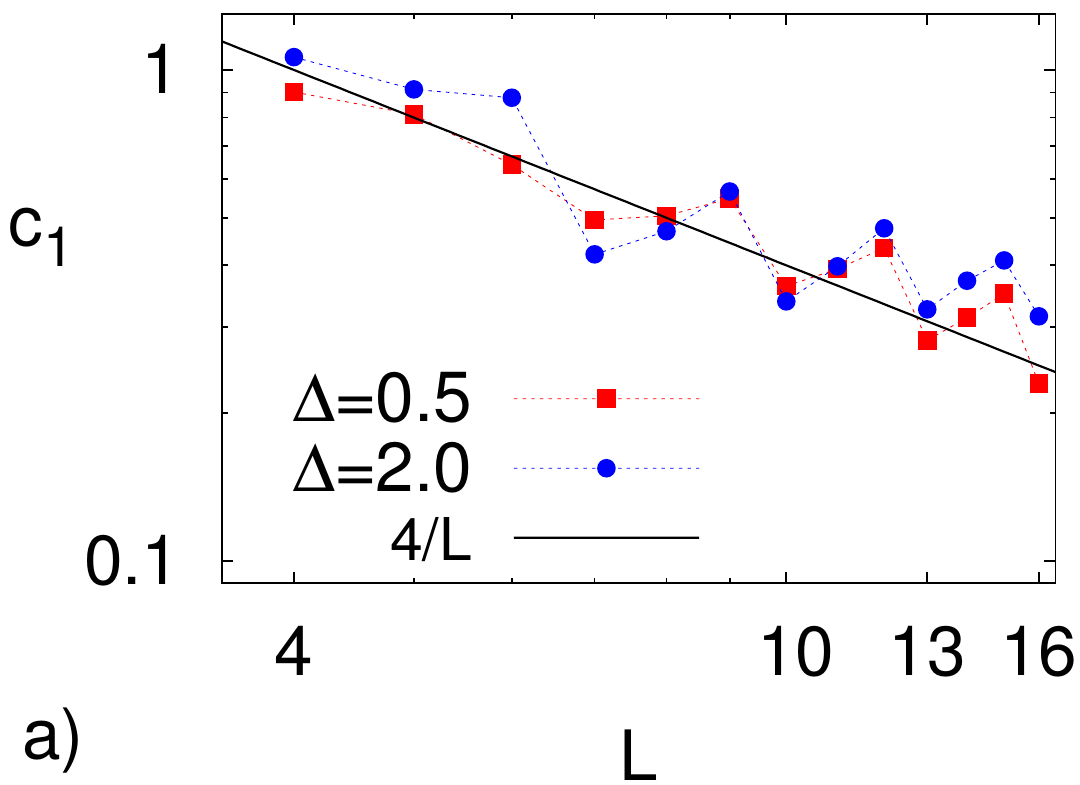}
\includegraphics[width=1.65in]{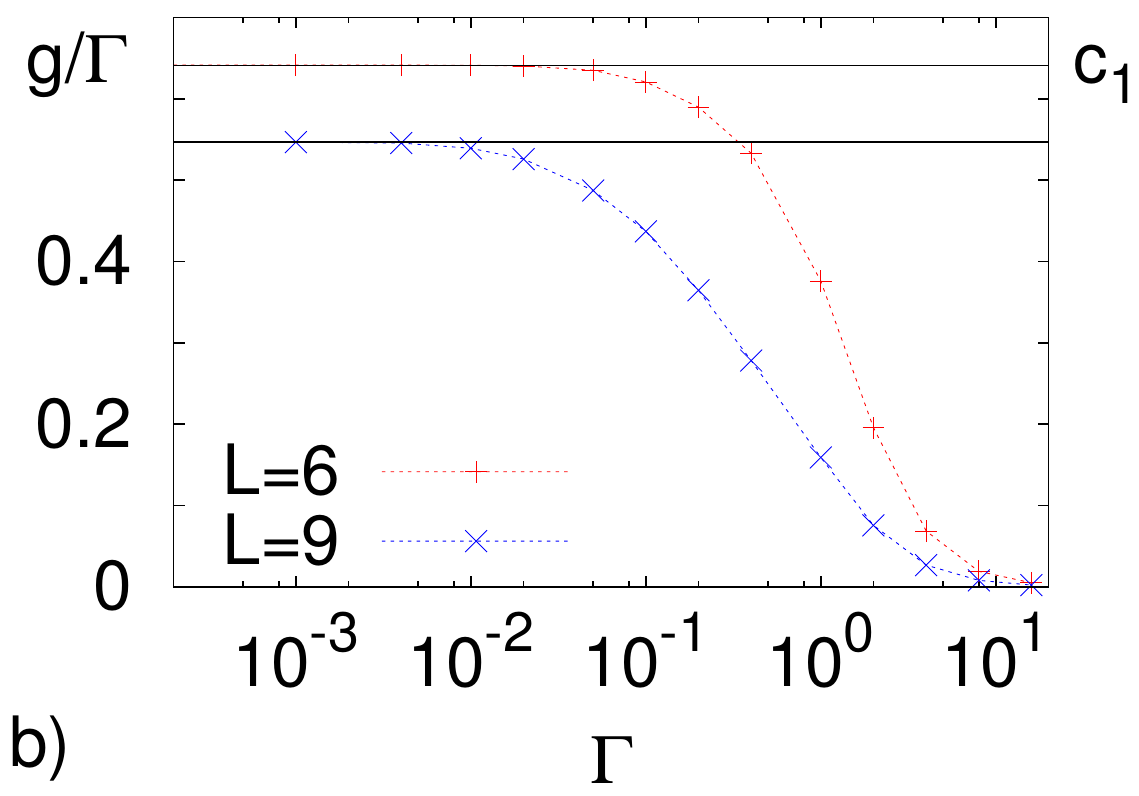}
}
\caption{(Color online) Staggered XXZ chain with weak boundary magnetization driving. (a) The largest non-zero eigenvalue of $R$, Eq.~(\ref{eq:muR}), determining the gap for small $\Gamma$, scales as $\sim 1/L$.  (b) Agreement between perturbative $c_1$ from (a) with the exact gap $g(\Gamma)$, all for $\Delta=0.5$. Convergence radius $\Gamma_{\rm c}$ decreases with $L$.}
\label{fig:staggmupert}
\end{figure}
\begin{figure}[t!]
\includegraphics[width=3.2in]{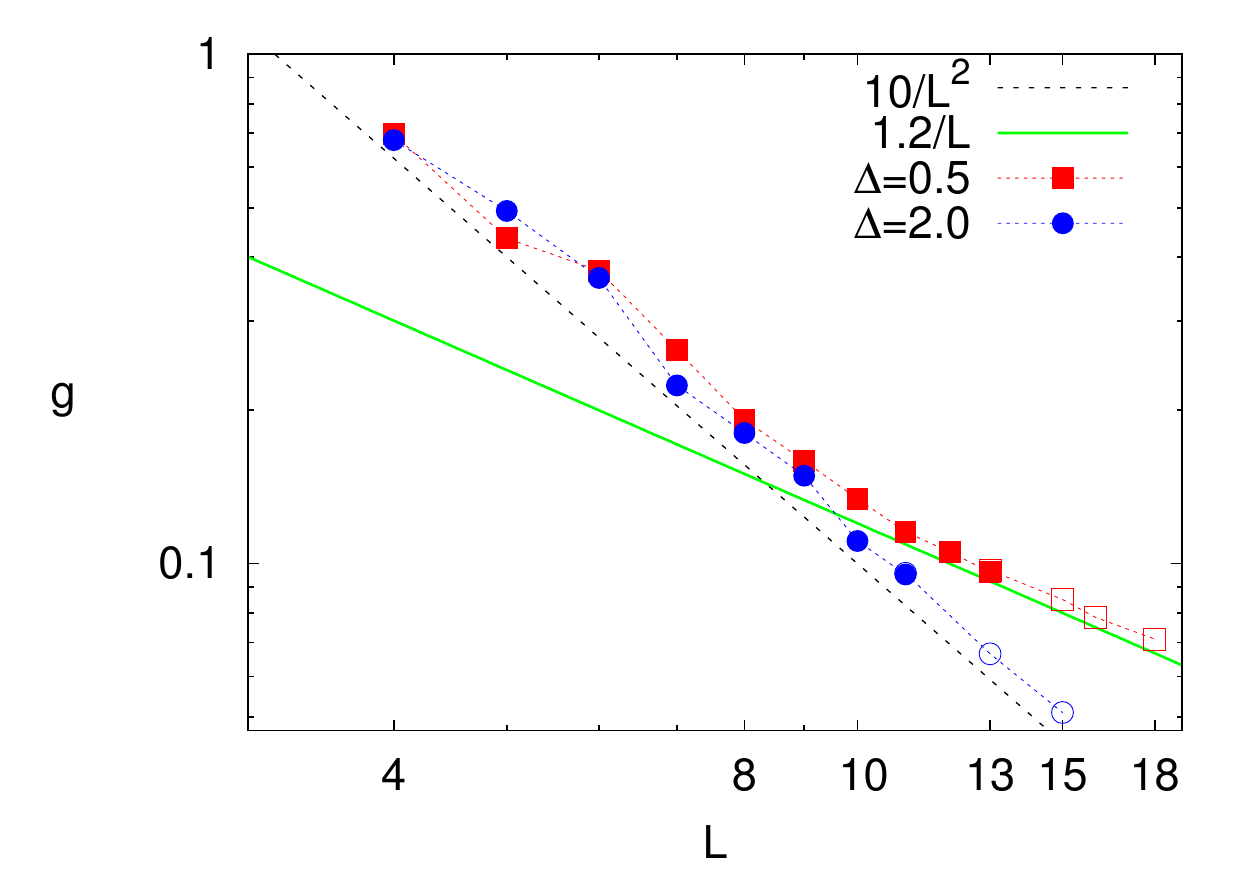}
\caption{(Color online) The gap $g$ for the chaotic XXZ model with staggered field (\ref{eq:XXZstagg}) and boundary magnetization driving (\ref{eq:mu}), $\mu=0.1, \Gamma=1$. Full symbols are exact diagonalization while empty are obtained from tDMRG. Straight lines suggest $\propto 1/L$ and $\propto 1/L^2$ asymptotic behavior.}
\label{fig:drivstaggXXZ}
\end{figure}
For small dissipation one can again use perturbation analysis in dissipation strength $\Gamma$. Everything is similar to the case of the XXZ model without staggered field. The perturbation matrix $R$ is given in Eq.~(\ref{eq:muR}). For the XXZ model with a staggered field eigenvalues of $R$ do dependent on $\mu$, however dependence of $c_1$ is very weak, with the correction at $\mu=0.1$ being less than $1 \%$. Therefore one could again use a simplified expression for $R$ given in Eq.~(\ref{eq:muRsimpl}). Data in Fig.~\ref{fig:staggmupert} show that $c_1$, and therefore also $g$ for small $\Gamma$, scales as $\sim 1/L$, with the range of validity (convergence radius $\Gamma_{\rm c}$) decreasing with system size $L$. Observe also that the effect of staggered field is much more pronounced than in the staggered XXZ with boundary dephasing (Fig.~\ref{fig:staggXXZpert}).

Going to non-small values of dissipation $\Gamma$, Fig.~\ref{fig:drivstaggXXZ} , the asymptotic scaling of the gap seem to be $\sim 1/L$ for $\Delta<1$, while for $\Delta>1$ it looks like $g \sim 1/L^2$, in both cases though convergence is less clear than in other models. For $\Delta<1$ the asymptotic gap certainly seems to be larger than $\sim 1/L^2$, showing that the Liouvillian gap for a boundary-only dissipation does not necessarily reflect diffusivity of the Hamiltonian.

\subsubsection{Tilted Ising with boundary dephasing}

\begin{figure}[t!]
\includegraphics[width=3.2in]{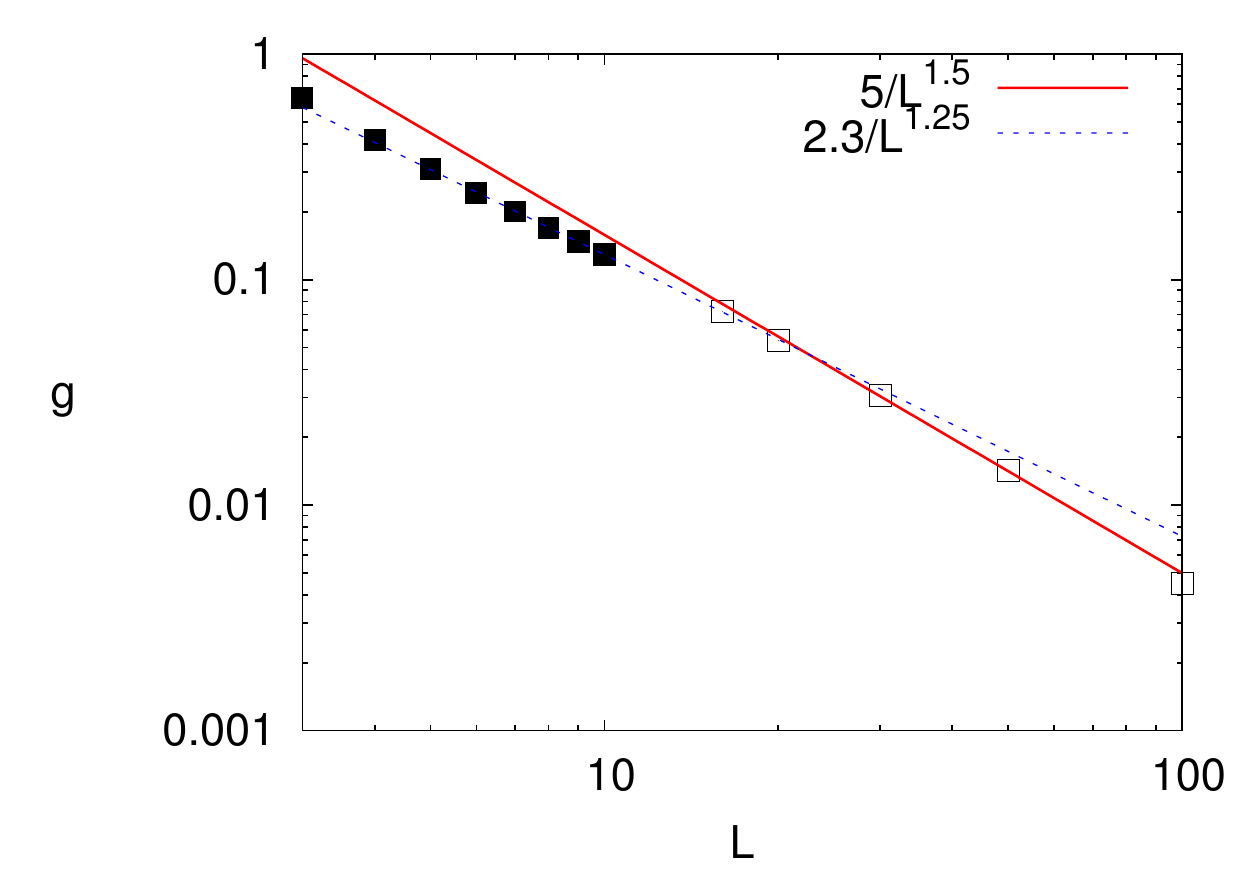}
\caption{(Color online) Gap $g$ for the tilted Ising model (\ref{eq:ti}) with boundary dephasing of strength $\gamma=1$. Points are numerical calculations: full squares for exact diagonalization, and empty squares for tDMRG.}
\label{fig:ti}
\end{figure}

The last model that we are going to study is again quantum chaotic one, but one that no longer conserves $z$. We take the Ising chain in a tilted magnetic field,
\begin{equation}
H=\sum_{j=1}^{L-1} -2\sz_j \sz_{j+1}+\sum_{j=1}^L b_x \sx_j+b_z \sz_j,
\label{eq:ti}
\end{equation}
which is quantum chaotic~\cite{Mejia07} for generic field direction (we use $b_x=3.375$ and $b_z=2$) and with diffusive energy transport~\cite{Mejia07,JSTAT09}. For small dephasing strength $\gamma$ the gap will again be equal to $g \approx \gamma c_1$, where $c_1$ is the largest non-zero eigenvalue of matrix $R$ (\ref{eq:boundaryR}). Numerically computed $c_1$ for sizes $L\le 15$ show scaling $c_1 \propto 1/L^\alpha$ with $\alpha \approx 0.85-1.0$ (data not shown).

We have also numerically calculated $g$ (Fig.~\ref{fig:ti}) for a system with boundary dephasing (\ref{eq:dephasing}) of strength $\gamma=1$. For smaller $L$ we used exact full diagonalization or the Arpack package, while for $L >10$ we used tDMRG~\cite{foot5}, inferring $g$ from the observed relaxation rate of total energy, the initial state being a pure N\' eel state $\ket{0101\ldots}\bra{0101\ldots}$. For this model we could get $g$ with tDMRG for significantly larger systems than for the other two chaotic models. We can see from Fig.~\ref{fig:ti} that the decay seems to be algebraic $1/L^\alpha$ with $\alpha$ being between $1-2$. While smaller $L$ seem to be described by $1/L^{1.25}$, for larger we get nicer fit with $1/L^{1.5}$. We in principle can not exclude the scaling becoming $\sim 1/L^2$ for still larger $L$, however, for the available data the scaling seems to be distinctively different than $\sim 1/L^2$.

\section{Summary and Conclusion} 

Let us briefly summarize our results and point to some interesting findings and open issues. We have studied open quantum spin chains with two types of sites at which dissipation acts. The first case was chains with bulk dissipation in which dissipation acts on all (or most) sites, while the second case was chains with boundary dissipation. Scaling of the gap for models with bulk dissipation is summarized in Table~\ref{tab:bulk} .
\begingroup
\squeezetable
\begin{table}[t!]
\begin{ruledtabular}
\begin{tabular}{rcrcrrcr}
 & \phantom{abc} & \multicolumn{6}{c}{{\bf Bulk dissipation}}\\
\multicolumn{1}{l}{Sector} & \phantom{abc} & XX+deph. & \phantom{abc} & \multicolumn{2}{c}{XXZ+deph.} & \phantom{abc} & XX+hopp. \\
\cmidrule(r){3-3} \cmidrule(r){5-6} \cmidrule(l){8-8}
  && && $\Delta<1$ & $\Delta>1$ &&\\
\midrule
\multicolumn{1}{l}{1-particle:}\\
$g(\gamma>\gamma_{\rm c})$ && $1/L^2$ && $1/L^2$ & $1/L^2$ && $1/L^0$\\
$g(\gamma<\gamma_{\rm c})$ && $1/L^0$ && $1/L^0$ & $1/L^0$ && \\
$\gamma_{\rm c}$ && $1/L$ && $1/L$ &$1/L$ \\
\multicolumn{1}{l}{Half-filling:}\\
$g(\gamma>\gamma_{\rm c})$ && $1/L^2$ && $1/L^2$ & $1/L^2$ && \\
$g(\gamma<\gamma_{\rm c})$ && $1/L^0$ && $1/L^0$ & $\approx 1/L^{0.8}$ && \\
$\gamma_{\rm c}$ && $1/L$ && $1/L^{1.2}$ & ${\rm e}^{-k L}$ &&
\end{tabular}
\end{ruledtabular}
\caption{Asymptotic scaling of the gap $g$ with system size $L$ for studied spin chains with bulk dissipation. Gap scaling for perturbatively weak dissipation, $\gamma < \gamma_{\rm c}$, as well as for non-small dissipation is listed. Behavior of $\gamma_{\rm c}$ with increasing $L$ is listed in two rows with heading ``$\gamma_{\rm c}$''. Approximate sign $\approx$ means that the scaling was inferred from small $L$ and that the observed scaling is perhaps not yet the asymptotic one.}
\label{tab:bulk}
\end{table}
\endgroup

\begingroup
\squeezetable
\begin{table*}[!t]
\begin{ruledtabular}
\begin{tabular}{rc r c rr c rr c rr c rr c r}
 & \phantom{abc} & \multicolumn{15}{c}{{\bf Boundary dissipation}}\\
\multicolumn{1}{l}{Sector} & \phantom{abc} & XX+deph. & \phantom{abc} & \multicolumn{2}{c}{XXZ+deph.} & \phantom{abc} & \multicolumn{2}{c}{XXZ+$\mu$.} & \phantom{abc} & \multicolumn{2}{c}{stagg.~XXZ+deph.} & \phantom{abc} & \multicolumn{2}{c}{stagg.~XXZ+$\mu$} & \phantom{abc} & tilted Ising+deph.  \\
\cmidrule(r){3-3} \cmidrule(r){5-6} \cmidrule(r){8-9} \cmidrule(r){11-12} \cmidrule(r){14-15} \cmidrule(l){17-17} 
  && && $\Delta<1$ & $\Delta>1$ && $\Delta<1$ & $\Delta>1$ && $\Delta<1$ & $\Delta>1$ && $\Delta<1$ & $\Delta>1$ &&  \\
\midrule
\multicolumn{1}{l}{1-particle:}\\
$g(\gamma>\gamma_{\rm c})$ && $1/L^3$ && $1/L^3$ & $1/L^3$ && \multicolumn{2}{c}{\multirow{2}{*}{not conserved}}  && $1/L^3$ & $1/L^3$ && \multicolumn{2}{c}{\multirow{2}{*}{not conserved}} && \multirow{2}{*}{not conserved}\\
$g(\gamma<\gamma_{\rm c})$ && $1/L^3$ && $1/L^3$ & $1/L^3$ && & && $1/L^3$ & $1/L^3$ &&\\
\multicolumn{1}{l}{Half-filling:}\\
$g(\gamma>\gamma_{\rm c})$ && $1/L^3$ && $1/L^3$ & ${\rm e}^{-\alpha L}$ && $1/L^3$ & $1/L^3$ && $1/L$ & ${\rm e}^{-\alpha L}$ && $\approx 1/L$ & $\approx 1/L^2$ && $\approx 1/L^{1.5}$\\
$g(\gamma<\gamma_{\rm c})$ && $1/L^3$ && $1/L^3$ & ${\rm e}^{-\alpha L}$ && $1/L^3$ & $1/L^3$ && $1/L$ & ${\rm e}^{-\alpha L}$ && $1/L$ & $\approx 1/L$ && $\approx 1/L^{0.9}$\\
$\gamma_{\rm c}$ or $\Gamma_{\rm c}$ && $L^0$ && \multicolumn{2}{c}{$L^0$} & & \multicolumn{2}{c}{$\Gamma_{\rm c} \sim L^0$} && $\gamma_{\rm c}\to 0$ & $\approx L^0$ && $\Gamma_{\rm c}\to 0$ & $\Gamma_{\rm c}\to 0$
\end{tabular}
\end{ruledtabular}
\caption{Asymptotic scaling of the gap $g$ with system size $L$ for studied spin chains with boundary dissipation. For systems that do not conserve particle number $r$ the global gap is listed under the ``half-filling'' heading. When known, we also list behavior with increasing system size of the convergence radius $\gamma_{\rm c}$ (or $\Gamma_{\rm c}$ for ``magnetization'' driving) of perturbation series in dissipation. ``Deph'' in the model description denotes dephasing dissipation, Eq.~(\ref{eq:dephasing}), ``$\mu$'' magnetization driving, Eq.~(\ref{eq:mu}).}
\label{tab:bound}
\end{table*}
\endgroup

Not surprisingly, because there are no fundamental limitations on the gap, one also finds all different scalings. Quite universally, the gap scales differently for small dissipation strength than for non-small dissipation (dissipation strength is in our cases mostly dephasing $\gamma$). Systems with bulk dissipation therefore typically undergo a (nonequilibrium) phase transition from a phase dominated by $H$ to a phase where dissipation is dominating. Such transition in the decay mode has been analyzed in detail for the XX chain with dephasing. In our models the critical dissipation strength $\gamma_{\rm c}$ at which this transition happens always goes to zero in the thermodynamic limit.

For boundary driven models the gap scaling is summarized in Table~\ref{tab:bound} . They all comply with a general bound prohibiting faster than $g \sim 1/L$ relaxation. Compared to bulk-dissipated cases, here the scaling seems the same for small and for non-small dissipation (except perhaps for the tilted Ising case, and the staggered XXZ model with magnetization driving; in both cases though finite-size effects could still be at play). In the 1-particle sector the gap is always $\sim 1/L^3$ due to essentially the solvability of the smallest nontrivial subspace (eventhough a model might be chaotic in larger invariant subspaces) and the longest eigenmodes having wavelength $\propto L$. 

A number of transitions in the scaling of $g$ can be identified. At each such transition there is a possible (nonequilibrium) phase transition that would be interesting to explore in more detail. Changing a bulk parameter, like the anisotropy $\Delta$ or the staggered field, can change the scaling (e.g., from algebraic to exponential, or from $\sim 1/L^3$ to $\sim 1/L$). More interestingly, the scaling can also change by changing a boundary dissipation only, that is, changing only terms that have a relative weight ${\cal O}(1/L)$ in the Liouvillian $\cL$. An example is the gapped XXZ chain for which the gap is exponentially small if dephasing is at the boundary, while it is algebraic for magnetization driving at the boundary. Such a transition is due to a symmetry breaking of otherwise protected subspace.

One finding that needs further exploration is a fast $g \sim 1/L$ relaxation in chaotic models. In the present work we studied only the spectral gap, without detailed discussion of the associated eigenvector properties. Of particular interest is locality of decay modes and with it connected relaxation of local observables. Namely, it has been observed in Lindblad equations as well as in classical systems~\cite{Schutz} that (certain) local observables can relax in a time that is smaller and scales differently than the global gap. For instance, in the XXX chain with Lindblad magnetization driving the gap scales as $g \sim 1/L^3$ whereas local magnetization and current relax as $\sim 1/L^{3/2}$~\cite{JSTAT11}.

\appendix
\section{Uniform-mixture steady state}
\label{app:A}
In all systems studied that conserve magnetization as well as the number of particles in the bra and ket, that is $z$ and $r$ (\ref{eq:const}), the Liouvillian eigenproblem has a block structure. In the $z=0$ sector one has a steady state in each $r-$particle sector, and that steady state is an equal mixture of projectors to all basis states (an ergodic diagonal state),
\begin{equation}
\rho=\frac{1}{{L\choose r}}\sum_j \ket{\psi_j}\bra{\psi_j},
\end{equation}
with $\ket{\psi_j}$ having $r$ spins in state $\ket{0}$ and $L-r$ in state $\ket{1}$. The two most interesting subspaces are the 1-particle and the $L/2$-particle (half-filling).

\subsection{One-particle sector}
The simplest case is $L=2$, for which the steady state is
\begin{equation}
\rho_2 = \frac{1}{2}(\ket{10}\bra{10}+\ket{01}\bra{01}).
\end{equation}
Taking a local operator basis composed of $\{ \ket{0}\bra{0},\ket{1}\bra{1},\ket{0}\bra{1},\ket{1}\bra{0} \}$ one can ``vectorize'' the operator $\rho$, writing it as a vector in a Hilbert space of operators. Doing that on $\rho_2$, it can be written as
\begin{equation}
\kket{\rho_2} = \frac{1}{\sqrt{2}}(\kket{10}+\kket{01}),
\end{equation}
where we use the notation $\kket{\bullet}$ to denote vectors in the space of operators, written in the operator basis $\{ \kket{0},\kket{1},\kket{2},\kket{3}\} \equiv \{ \ket{0}\bra{0},\ket{1}\bra{1},\ket{0}\bra{1},\ket{1}\bra{0}\}$. The steady state in the $1-$particle sector on $L$ spins, $\rho_L$, is simply
\begin{equation}
\kket{\rho_L} =\frac{1}{\sqrt{L}}(\kket{10\ldots 0}+\kket{010\ldots 0}+\cdots+\kket{0\ldots 01}),
\end{equation}
which is the so-called $W$-state. One can immediately see that for a bipartite cut after $p$ spins the Schmidt decomposition is of rank $2$ with the two eigenvalues (squares of Schmidt coefficients) being $1-\frac{1}{L}$ and $\frac{1}{L}$, irrespective of $p$. In the operator space such a steady state is therefore of finite rank $2$ (it is weakly entangled) regardless of the system size $L$.

\subsection{Half-filling sector}
The half-filling sector is composed of states with half of the spins pointing up and half pointing down. Total magnetization is therefore zero. Starting again with a simple example for $L=4$, we have the steady state
\begin{eqnarray}
\rho_4&=&\frac{1}{6}(\ket{0011}\bra{0011}+\ket{0101}\bra{0101}+\ket{1001}\bra{1001}+\nonumber \\
&& \!\!\!\!\!\! + \ket{0110}\bra{0110}+\ket{1010}\bra{1010}+\ket{1100}\bra{1100}),
\end{eqnarray}
or, written as a vector,
\begin{eqnarray}
\kket{\rho_4}=\frac{1}{\sqrt{6}}&\big( &\kket{0011}+\kket{0101}+\kket{1001}+\kket{0110}+ \nonumber \\
&& +\kket{1010}+\kket{1100}\big),
\end{eqnarray}
where $\kket{\rho_L}$ is a uniform superposition of all ${L \choose L/2}$ basis states with zero magnetization (for simplicity, we assume even $L$; for odd $L$ and the largest sector one has to replace $L/2$ with $(L+1)/2$). Regarding (operator) Schmidt decomposition, for a cut after $p=1$ sites we see that the Schmidt rank is $2$ with both eigenvalues being $\frac{1}{2}$. For $p=2$, i.e., a cut of maximal size for $\rho_4$, we have decomposition
\begin{eqnarray}
\kket{\rho_4} &\sim& \kket{00}\kket{11}+\kket{11}\kket{00}+\\
&&+\sqrt{4}\frac{1}{\sqrt{2}}(\kket{01}+\kket{10})\frac{1}{\sqrt{2}}(\kket{01}+\kket{10}) \nonumber,
\end{eqnarray}
and therefore the state is of rank $3$ with eigenvalues being $\frac{1}{6},\frac{1}{6},\frac{4}{6}$. We can see that the eigenvalue prefactors ($1,1,4$) are actually of a combinatorial nature, resulting from the number of combinations of distributing $k$ ones on $p$ sites, e.g., $1={2 \choose 0}{2 \choose 2}$, $1={2 \choose 2}{2 \choose 0}$, and $4={2 \choose 1}{2 \choose 1}$. Generalizing to a bipartite cut of $\kket{\rho_L}$ after $p$ sites, one has $p+1$ nonzero Schmidt coefficients with the eigenvalues being ${p \choose k}{L-p \choose L/2-k}/{L \choose L/2},\quad k=0,\ldots,p$. We can see that the largest Schmidt rank is for a half-cut, and is equal to $L/2+1$. Operator Schmidt rank of the steady state in the half-filling sector grows linearly with the system size $L$. We remind that a constant~\cite{JPA10} or a linear~\cite{Prosen11} Schmidt rank is a sign of an exact solvability of a steady state. However, our results show that the exact solvability of the Lindblad steady state in general does not tell us anything about solvability of (closest) decay modes or the behavior of the gap.


\begin{thebibliography}{99}
\bibitem{foot1} Ref.~\cite{Gaspard:05} uses periodic boundary conditions, facilitating a solution in Fourier space. We prefer to keep open boundary conditions with a view of later-studied boundary-driven systems, where the first and the last sites are typically not coupled and the translational invariance is anyway lost due to localized dissipation.
\bibitem{foot2} The mapping from $\cL$ to the ladder is such~\cite{PRE14} that ladder states from the 1-particle subspace have e.g. $L-1$ particles in $\ket{0}$ and one in $\ket{1}$ in the upper leg, and $L-1$ in $\ket{1}$ and one in $\ket{0}$ in the lower leg.
\bibitem{foot3} For system sizes upto around $L \approx 10$ we used full exact diagonalization of $\cL$ in the respective half-filling subspace, for $L$ upto $15$ the Lanczos method (ARNOLDI package), and for still larger $L$ the tDMRG method, see e.g.~\cite{NJP10} for further details and references about the tDMRG method.
\bibitem{foot5} We have checked that tDMRG gave the same results as exact diagonalization for $L<10$.
\end{thebibliography}
\end{document}